\documentclass[runningheads]{llncs}

\usepackage{booktabs}   %% For formal tables:
\usepackage{subcaption} %% For complex figures with subfigures/subcaptions
\usepackage{latexsym}
\usepackage{amssymb}
\usepackage{amsmath}
\usepackage{stmaryrd}
\usepackage[all]{xy}
\usepackage{graphicx}
\usepackage{hyperref}
\usepackage{multicol}
\usepackage{bm}
\usepackage{cancel}

\newcommand{\cons}{\!:\!}

\usepackage{tikz}
\usetikzlibrary{calc,arrows,shadows,positioning}

\newcommand{\ignore}[1]{}

\newcommand{\arro}[1]{\xrightarrow{#1}} 
\newcommand{\boo}{\rightarrowtail} 
\newcommand{\hoo}{\hookrightarrow} 

\newcommand{\rolldel}{\mathsf{rolldel}} 
\newcommand{\comp}{\:|\:}

\newcommand{\rlh}{\rightleftharpoons}
\newcommand{\lh}{\leftharpoondown}
\newcommand{\rh}{\rightharpoonup}

\newcommand{\sql}{\mbox{$\lfloor\hspace{-.5ex}\lfloor$}}
\newcommand{\sqr}{\mbox{$\rfloor\hspace{-.4ex}\rfloor$}}

\newcommand{\sqll}{\mbox{$\lceil\hspace{-.5ex}\lceil$}}
\newcommand{\sqrr}{\mbox{$\rceil\hspace{-.4ex}\rceil$}}

\usepackage{xcolor}

\newcommand{\red}[1]{{\color{red} #1}}

\newcommand{\blue}[1]{{\color{blue} #1}}

\newcommand{\id}{id}
\newcommand{\cC}{{\mathcal{C}}}

\newcommand{\nil}{[]}

\def \tuple#1{\langle #1 \rangle}
\def \ttuple#1{\langle\!\langle #1 \rangle\!\rangle}

\def \tupleb#1{(\!( #1 )\!)}

\def \tuplefb#1{(\!( #1 )\!)}

\long\def\comment#1{}

\begin{document}

\title{%
An Asynchronous Scheme for Rollback Recovery in 
Message-Passing Concurrent Programming Languages%
\thanks{
This work has been partially supported by 
grant PID2019-104735RB-C41
funded by MCIN/AEI/ 10.13039/501100011033, by French ANR project DCore ANR-18-CE25-0007, and by  
\emph{Generalitat Valenciana} under grant CIPROM/2022/6 
(FassLow).
}
}

\titlerunning{An Asynchronous Scheme for Rollback Recovery in 
Message-Passing Concurrent Programming Languages}

\author{Germ\'an Vidal}

\authorrunning{G. Vidal}

\institute{
  VRAIN, Universitat Polit\`ecnica de Val\`encia\\
\email{gvidal@dsic.upv.es}
}

\maketitle

%\title{An Asynchronous Scheme for Rollback Recovery in 
%Message-Passing Concurrent Programming Languages}%
%\titlenote{This work has been partially supported by 
%grant PID2019-104735RB-C41
%funded by MCIN/AEI/ 10.13039/501100011033, 
%by French ANR project DCore ANR-18-CE25-0007, and by  
%\emph{Generalitat Valenciana} under grant CIPROM/2022/6 
%(FassLow).}
%%\subtitle{Extended Abstract}
%%\subtitlenote{The full version of the author's guide is available as
%%  \texttt{acmart.pdf} document}
%  
%%\renewcommand{\shorttitle}{SIG Proceedings Paper in LaTeX Format}
%
%
%%\author{Author1 Author2 Author2 \ldots}
%\author{Germ\'an Vidal}
%%\authornote{This work has been partially supported by 
%%grant PID2019-104735RB-C41
%%funded by MCIN/AEI/ 10.13039/501100011033, 
%%by French ANR project DCore ANR-18-CE25-0007, and by  
%%\emph{Generalitat Valenciana} under grant CIPROM/2022/6 
%%(FassLow).}
%\orcid{0000-0002-1857-6951}
%\affiliation{%
%  \institution{Universitat Polit\`ecnica de Val\`encia}
%  \streetaddress{Camino de Vera, S/N}
%  \city{Valencia} 
%%  %\state{} 
%  \country{Spain}
%  \postcode{46022}  
%}
%\email{gvidal@dsic.upv.es}

\begin{abstract}
	Rollback recovery strategies are well-known in 
	concurrent and distributed systems. In this context,
	recovering from unexpected failures is even more 
	relevant given the non-deterministic nature of 
	execution, which means that it is practically 
	impossible to foresee all possible process interactions.
	
	In this work, we consider a message-passing concurrent
	programming language where processes interact through
	message sending and receiving, but shared memory
	is not allowed. In this context, we design a
	checkpoint-based
	rollback recovery strategy that does not 
	need a central coordination. 
	For this purpose, we extend the language
	with three new operators: $\mathsf{check}$,
	$\mathsf{commit}$, and $\mathsf{rollback}$. 
	Furthermore, our approach
	is purely asynchronous, which is an essential 
	ingredient to developing a source-to-source program 
	instrumentation implementing a rollback recovery
	strategy. % for achieving fault tolerance.
	\\[2ex]
    \emph{{\copyright} Germ\'an Vidal. This is the author's version 
    of the work. It is posted here for your personal use. 
    Not for redistribution. The definitive version was published 
    in the Proceedings of the 39th ACM/SIGAPP Symposium on Applied 
    Computing (SAC '24), April 8--12, 2024, Avila, Spain, ACM.\\ 
    \url{https://doi.org/10.1145/3605098.3636051}.}	
%\footnote{This is an abstract footnote}
\end{abstract}

%\begin{CCSXML}
%<ccs2012>
%   <concept>
%       <concept_id>10003752.10003753.10003761</concept_id>
%       <concept_desc>Theory of computation~Concurrency</concept_desc>
%       <concept_significance>500</concept_significance>
%       </concept>
%   <concept>
%       <concept_id>10011007.10011074.10011092.10011691</concept_id>
%       <concept_desc>Software and its engineering~Error handling and recovery</concept_desc>
%       <concept_significance>500</concept_significance>
%       </concept>
%   <concept>
%       <concept_id>10011007.10011006</concept_id>
%       <concept_desc>Software and its engineering~Software notations and tools</concept_desc>
%       <concept_significance>300</concept_significance>
%       </concept>
%\end{CCSXML}
%
%\ccsdesc[500]{Theory of computation~Concurrency}
%\ccsdesc[500]{Software and its engineering~Error handling and recovery}
%\ccsdesc[300]{Software and its engineering~Software notations and tools}


%\maketitle

%%%%%%%%%%%%%%%%%%%%%%%%%%%%%%%%%%%%%%%%%%%%%
\section{Introduction} \label{sec:intro}

Some popular approaches to rollback recovery in message
passing systems can be found in the survey by 
Elnozahy \emph{et al} \cite{EAWJ02}. 
Most of these approaches are based on so called
\emph{checkpointing}, where processes save 
their state periodically 
so that, upon a failure, the system can use the saved
states---called checkpoints---to recover a previous
but consistent state of the system. 

In contrast to \cite{EAWJ02}, which is focused on
\emph{transparent} approaches to rollback recovery,
our proposal is oriented to extending a programming
language with \emph{explicit} rollback recovery operators.
In particular, we consider the following three
basic operators:
\begin{itemize}
\item $\mathsf{check}()$: it saves the current state
of the process (a checkpoint) and returns a unique
identifier, e.g., $\tau$.
\item $\mathsf{commit}(\tau)$: this call commits
a checkpoint $\tau$, i.e., the computation performed
since the call to $\mathsf{check}()$ is considered
definitive and the state saved in checkpoint $\tau$ 
is discarded.
\item $\mathsf{rollback}(\tau)$: this call is used to
recover a saved state (the one associated to checkpoint 
$\tau$).  
\end{itemize}
We consider in this work a typical message-passing
(asynchronous) concurrent programming language like,
e.g., (a subset of) Erlang \cite{erlang}. 
The considered language
mostly follows the actor model \cite{HBS73}, where
a running application consists of a number of processes
(or actors) that can only communicate through message
sending and receiving, but shared memory is not allowed.
Furthermore, we consider that processes can be 
dynamically spawned at run-time, in contrast to 
session-based programming based on (multiparty) 
session types \cite{HYC08} where the number of 
partners is typically fixed.

As is common, when a process rolls back to a particular
checkpoint, we require the entire system to be 
\emph{causally consistent}, i.e., no message can be
received if---after the rollback---it has not been sent,
or no process may exist if it has not been spawned.
This notion of \emph{causality} follows the
well-known Lamport's ``happened before'' relation 
\cite{Lam78},which says that action $a$
happened before action $b$ if
\begin{itemize}
\item both actions are performed by the same process
and $a$ precedes $b$,
\item action $a$ is the sending of a message and 
action $b$ is the receiving of this message, or
\item action $a$ is the spawning of a new process $p$
and action $b$ is any action performed by process $p$.
\end{itemize}
Hence, in order to have a causally consistent
rollback recovery strategy,
whenever a rollback operator is executed, one should
not only recover the corresponding previous state of 
this process, but possibly also propagate the
rollback operation to other processes. 

Extending the language with explicit operators for
rollback recovery can be useful in a number of 
contexts. For example, they can be used to improve
an ordinary ``try\_catch'' statement so that a
rollback is used to undo the actions performed so far 
whenever an exception is raised, thus
avoiding inconsistent states.
In general, this operators can be used to enforce
fault tolerance by allowing the user to define a
sort of \emph{transactions} so that either all of
them are performed or none (see, e.g., the combination
of message-passing concurrency and software
transactional memory in \cite{SKM17}).

%%%%%%%%%%%%%%%%%%%%%%%%%%%%%%%%%%%%%%%%%%%%%
\section{An Asynchronous 
Message-Passing Concurrent Language} \label{sec:lang}

In this section, we present the essentials of a simple
message-passing concurrent language where
processes can (dynamically) spawn new processes 
and can (only) interact through 
message sending and receiving (i.e., there is 
no shared memory). This is the case, e.g., of the
functional and concurrent language Erlang \cite{erlang},
which can be seen as a materialization of the 
\emph{actor model} \cite{HBS73}. 

Although we are not going to formally introduce the syntax
and semantics of the considered subset of Erlang 
(which can be found elsewhere, 
e.g., in \cite{LNPV18jlamp,LM20})
%\cite[Appendices A \& B]{GV21tr}),
let us illustrate it with a simple example:

\begin{figure}[t]
\begin{verbatim}
  init() -> S = spawn(fun() -> bank(100) end),
            spawn(fun() -> client(S) end).          
  bank(B) -> receive
               {C,get} -> 
                   C ! B, bank(B);
               {C,withdraw,N} -> 
                   try 
                      C ! ack,
                      ...  // some safety checks
                      C ! ok, bank(NB)
                   catch
                      _:_ -> bank(B)
                   end
             end.
client(S) -> S ! {self(),get},
             receive
                Amount -> ...
             end,
             S ! {self(),withdraw,50}, ...
\end{verbatim}
\caption{Example program (bank account server)} \label{fig:program}
\end{figure}

\begin{example} \label{ex:one}
  Consider the Erlang program 
  shown in Figure~\ref{fig:program},
  where we have a bank account server and a single client
  that performs a couple of operations.
Execution starts with a single process that calls function
\verb$init/0$,\footnote{As in Erlang, we denote
function symbols with $f/n$ where $n$ is the arity of function~$f$.
Moreover, variables start with an uppercase letter.} 
This process then spawns two new processes using the
predefined function \verb$spawn/1$: the ``bank account server''
and the ``client''. The argument of \verb$spawn$ contains the
function that should execute the new process (\verb$bank(100)$
and \verb$client(S)$, respectively).
Function \verb$spawn/1$ returns the pid (for \emph{p}rocess
\emph{id}entifier) of the spawned process, a fresh identifier
that uniquely identifies each running process.
Variable \verb$S$ is thus bound to the pid of the bank account
server.

The server is basically an endless loop
which is waiting for client requests. A \verb$receive$
statement is used for this purpose. In particular, this
receive statement accepts two types of messages:
\begin{itemize}
\item \verb${C,get}$, where variable \verb$C$ is the pid of 
the client, and \verb$get$ is a constant (called \emph{atom}
in Erlang);
\item \verb${C,withdraw,N}$, where variable 
\verb$C$ is the pid of 
the client, \verb$withdraw$ is a constant, and \verb$N$
is the amount to be withdrawn from the account.
\end{itemize}
Both requests include the pid of the client in order to
get a reply from the server. 
The client process only performs two operations.
First, it sends a request to the server
to get the current balance, where message sending is denoted
with a statement of the form \verb$target_pid ! message$.  
Then, it waits for an answer,
which will eventually bind variable \verb$Amount$ to
this balance.\footnote{Here, we are simulating a synchronous
communication between the client and the server by
sending the client's own pid (using the built-in function
{\tt self/0}),
and having a receive statement after the message
sending, a common pattern in Erlang.}
Then, after performing some operations (not shown),
it sends a second request to the server to
withdraw \$50. We omit the following operations to keep the 
example as simple as possible.
%\end{enumerate}

The \emph{state} of the bank account server (the current
balance) is stored in 
the argument of function \verb$bank/1$ (initiliazed to 
\$100 when the process was spawned). Then, depending on
the request, the server proceeds as follows:
\begin{itemize}
\item For a request to get the current balance, 
a message is sent back to the client: \verb$C ! B$,
and a recursive call \verb$bank(B)$ is performed
 to execute the receive statement again.
\item For a withdrawal request,
the server sends an ack to the client, then 
performs some safety checks (not shown) and either 
sends \verb$ok$ back to the client and updates
the balance to \verb$NB$,  
or cancels the
operation and does a recursive call \verb$bank(B)$ with
the old balance (if an exception is raised during the
safety checks). 
We omit part of the code for simplicity.
\end{itemize}

\begin{figure}[t]
\centering
\begin{tikzpicture}
\draw[->,dashed,thick] (0,3.5) node[above]{\tt init} -- (0,0);
\draw[->,dashed,thick] (3,3.5) node[above]{\tt bank} -- (3,0);
\draw[->,dashed,thick] (6,3.5) node[above]{\tt client} -- (6,0);

\draw[->,dotted] (0,3.2) node[left]{$\mathsf{spawn}$} -- (3,3.1);

\draw[->,dotted] (0,2.8) node[left]{$\mathsf{spawn}$} -- (6,2.6);

\draw[->] (6,2.3) node[right]{$\mathsf{send}$} -- (3,2.1)
node[midway,above]{\tt get} node[left]{$\mathsf{rec}$};

\draw[->] (3,1.8) node[left]{$\mathsf{send}$} -- (6,1.6)
node[midway,above]{$B$} node[right]{$\mathsf{rec}$};

\draw[->] (6,1.3) node[right]{$\mathsf{send}$} -- (3,1.1)
node[midway,above]{\tt withdraw} node[left]{$\mathsf{rec}$};

\draw[->] (3,0.8) node[left]{$\mathsf{send}$} -- (6,0.6)
node[midway,above]{\tt ack} node[right]{$\mathsf{rec}$};

\draw[->] (3,0.4) node[left]{$\mathsf{send}$} -- (6,0.2)
node[midway,above]{\tt ok} node[right]{$\mathsf{rec}$};
\end{tikzpicture}
\caption{Graphical representation of the execution in
Example~\ref{ex:one} (time flows from top to bottom)}
\label{fig:ex-standard}
\end{figure}

\noindent
A graphical representation of the program's 
execution---assuming the safety checks are passed---is 
shown in Figure~\ref{fig:ex-standard}.
\end{example}
In the remainder of the paper, 
we will ignore the \emph{sequential} component of
the language and will focus on its
concurrent actions: process spawning, message sending,
and message receiving. Some features of the considered
language follows:
\begin{itemize}
\item processes can be dynamically spawned
at run time;
\item message-passing is asynchronous;\footnote{Nevertheless,
synchronous communication can be simulated using a combination
of message sending and receiving, as seen in Example~\ref{ex:one}.}
\item message receiving suspends the execution until a matching
message reaches the process;
\item messages can be delivered to a process at any point in
time and stored in a local \emph{mailbox} (a queue), but they will not 
be processed until a receive statement is executed (if any).
\end{itemize}
We let $s,s',\ldots$ denote \emph{states}, typically 
including an environment and an expression (or statement) to be 
evaluated. The structure of states is not relevant for the
purpose of this paper, though. 

\begin{definition}[process configuration]
  A process \emph{configuration} is denoted by a tuple of 
  the form $\tuple{p,s}$, where $p$ is the pid of the process
  and $s$ is its current state.
\end{definition}

\begin{definition}[message]
  A \emph{message} has the form $(p,p',v)$, where $p$ is the pid 
  of the sender, $p'$ that of the receiver, and $v$ is a 
  value.\footnote{We note that the pid of the sender is not really
  needed when the order of messages is not relevant. Nevertheless,
  we keep the current format for compatibility with other, related
  definitions (e.g., \cite{LPV21}).}
\end{definition}
A \emph{system} is either a process configuration, a message,
or the parallel composition of two systems $S_1\comp S_2$,
where ``$\comp$'' is commutative and
associative.
We borrow the idea of using \emph{floating}
messages from \cite{LSZ19} (in contrast to using a \emph{global} 
mailbox as in \cite{LNPV18jlamp}).

A floating message represents a message that has been
already sent but not yet delivered (i.e., the message is on the
network). Furthermore, process mailboxes
are abstracted away for simplicity, thus a floating message
can also represent a message that is already
stored in a process mailbox but is not yet consumed. As in
Erlang, we assume that the order of 
messages sent  directly from process $p$ to process $p'$ is 
preserved when they are  
all delivered. % (see~\cite[Section~10.8]{ErlangFAQ}).
We do not formalize this constraint 
for simplicity, but could easily be ensured by introducing
triples of the form $(p,p',vs)$ where $vs$ is a queue of messages
instead of a single message.

As in \cite{LNPV18jlamp,LPV21},
the semantics of the language is defined in a modular way, so that the
labeled transition relations $\arro{}$ and $\boo$ model the evaluation
of expressions (or statements) and the evaluation of 
systems, respectively. 

\begin{figure}[t]
\centering
$
  \begin{array}{r@{~~}c}
    (\mathit{Seq}) & {\displaystyle
      \frac{s \arro{\mathsf{seq}} s' 
      }{\tuple{p,s} 
        \boo \tuple{p,s'}}
      }\\[3ex]

    (\mathit{Send}) & {\displaystyle
      \frac{s \arro{\mathsf{send}(p',v)} s' 
      }{\tuple{p,s} 
        \boo
         (p,p',v) \comp \tuple{p,s'}}
      }\\[3ex]

      (\mathit{Receive}) & {\displaystyle
        \frac{s \arro{\mathsf{rec}(\kappa,cs)}
          s'~~\mbox{and}~~ \mathsf{matchrec}(cs,v) = cs_i}
          {(p',p,v) \comp
           \tuple{p,s} \boo
           \tuple{p,s'[\kappa\leftarrow cs_i])}}
      }\\[3ex]  

      (\mathit{Spawn}) & {\displaystyle
        \frac{s \arro{\mathsf{spawn}(\kappa,s_0)} 
          s'~~\mbox{and}~~ p'~\mbox{is a fresh pid}}
          {\tuple{p,s} \boo
           \tuple{p,s'[\kappa\leftarrow p']}\comp \tuple{p',s_0}}
      }\\[3ex]

    (\mathit{Par}) & {\displaystyle
      \frac{S_1 \boo S'_1~~\mbox{and}~~\id(S'_1)\cap \id(S_2)=\emptyset}{S_1  \comp S_2 
         \boo S'_1 \comp S_2 
        }
      }      
  \end{array}
$
\caption{Standard semantics} \label{fig:standard-semantics}
\end{figure}

We skip the definition of the local semantics 
($\to$) since it is not necessary for our developments; 
we refer the interested reader to \cite{LNPV18jlamp}. %,GV21tr}.
As for the rules of the operational semantics
that define the reduction of systems, we follow the
recent formulation in \cite{Vid23facs}.
The transition rules are shown 
in Figure~\ref{fig:standard-semantics}:
\begin{itemize}
\item Sequential, local steps are dealt with rule \emph{Seq}, which
propagates the reduction from the local level to the system
level.

\item Rule \emph{Send} applies when the local evaluation requires
sending a message as a side effect. The local step
$s \arro{\mathsf{send}(p',v)} s'$ is labeled with the information
that must flow from the local level to the system level:
the pid of the target
process, $p'$, and the message value, $v$. 
The system rule then adds a new message of the form $(p,p',v)$ 
to the system.

\item In order to receive a message, the situation is somehow different.
Here, we need some information to flow both from the local level
to the system level (the clauses $cs$ of the receive statement)
and vice versa (the selected clause, $cs_i$, if any).
For this purpose, in rule \emph{Receive}, the label of the local step
includes a special variable $\kappa$ ---a sort of \emph{future}--- 
that denotes the position of the receive expression within state $s$. 
The rule then checks if there is
a floating message $v$ addressed to process $p$
that matches one of the constraints in $cs$. This is done by the
auxiliary function $\mathsf{matchrec}$, which returns the
selected clause $cs_i$ of the receive statement in case of a match
(the details are not relevant here).
Then, the reduction proceeds by binding $\kappa$ in $s'$ with the
selected clause $cs_i$, which we denote by $s'[\kappa\leftarrow cs_i]$. 

\item Rule \emph{Spawn} also requires a bidirectional flow of
information. Here, the label of the local step includes
the future $\kappa$ and the state of the new
process $s_0$. It then produces a fresh pid,
%for the new process, 
$p'$, adds the new process $\tuple{p',s_0}$ to
the system, and updates the state $s'$ by binding $\kappa$
to $p'$ (since $\mathsf{spawn}$ reduces to the pid of the new
process), which we denote by $s'[\kappa\leftarrow p']$.

\item Finally, rule \emph{Par} is  used to lift an evaluation
step to a larger system. The auxiliary function
$\id$ takes a system $S$ and returns the set of pids in $S$,
in order to ensure that new pids are indeed fresh in the complete
system.
\end{itemize}
We let $\boo^\ast$ denote the 
transitive and reflexive closure of $\boo$. Given
systems $S_0,S_n$, $S_0 \boo^\ast S_n$ denotes
a \emph{derivation} under the standard semantics.
An \emph{initial} system has the form $\tuple{p,s_0}$, 
i.e., it contains a single process. 
A system $S'$ is \emph{reachable} if there exists a derivation
$S \boo^\ast S'$ such that $S$ is an initial system. 
A derivation $S \boo^\ast S'$ is \emph{well-defined} under the
standard semantics if $S$ is a reachable system.

As mentioned before, in this work we focus on the
concurrent actions of processes. For this purpose, in the
examples, we describe the actions of a process as a sequential
stream of concurrent actions, ignoring all other details
and hiding the structure of the underlying code.
In particular, we consider the following actions:
\begin{itemize}
\item \blue{$p\!\leftarrow\! \mathsf{spawn}()$}, for process spawning, 
where $p$ is the (fresh) pid of the new process (returned 
by the call to $\mathsf{spawn}$);
\item \blue{$\mathsf{send}(p,v)$}, for sending a message, where
$p$ is the pid of the target process and $v$ the message value;
\item \blue{$\mathsf{rec}(v)$}, for receiving message $v$.
\end{itemize}

\begin{example} \label{ex:two}
  For instance, the execution of Example~\ref{ex:one} as shown
  in Figure~\ref{fig:ex-standard} can be represented as follows:
	\blue{\[
	\begin{array}{l@{~~~~~~~~~~~~}l@{~~~~~~~~~~~~}l}
	 \mbox{\tt init} & \mbox{\tt bank} & \mbox{\tt client} \\\hline
	 {\tt bank}\!\leftarrow\!\mathsf{spawn}()
	 & \mathsf{rec}({\tt get})
	 & \mathsf{send}({\tt bank},{\tt get}) \\
	 
	 {\tt client} \!\leftarrow\!\mathsf{spawn}()
	 & \mathsf{send}({\tt client},B)
	 & \mathsf{rec}(B) \\

	 & \mathsf{rec}( {\tt withdraw})
	 & \mathsf{send}({\tt bank},{\tt withdraw}) \\

	 & \mathsf{send}({\tt client},{\tt ack})
	 &  \mathsf{rec}({\tt ack})\\

	 & \mathsf{send}({\tt client},{\tt ok})
	 &  \mathsf{rec}({\tt ok})\\

	\end{array}
	\]}
\end{example}

%%%%%%%%%%%%%%%%%%%%%%%%%%%%%%%%%%%%%%
\section{Operators for Checkpoint-Based Rollback Recovery}

Now, we present three new explicit
operators for checkpoint-based rollback recovery
in our message-passing concurrent language:
\begin{itemize}
\item $\mathsf{check}()$ introduces a \emph{checkpoint} for the 
current process. Its reduction %of \textsf{check}
returns a fresh identifier, $\tau$, associated to the checkpoint.
%\footnote{Note that nested checkpoints are possible.}
As a side-effect, the current state is saved.

\item $\mathsf{commit}(\tau)$ can then 
be used to discard the state saved in checkpoint $\tau$. 

\item Finally, $\mathsf{rollback}(\tau)$ recovers the
state saved in checkpoint $\tau$, possibly following a 
different execution path.
Graphically,
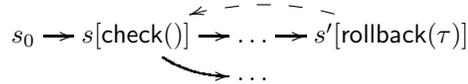
\[
\xymatrix@C=.4cm@R=.1cm{
s_0 \ar[r] 
& s[\mathsf{check}()] \ar[r] \ar@/_.5pc/[dr] & \ldots \ar[r] 
& s'[\mathsf{rollback}(\tau)] \ar@{-->}@/_1pc/[ll] \\
& & \ldots \\
}
\]
where $s[t]$ denotes an arbitrary state 
whose next expression to be reduced is $t$.
\end{itemize}
The reduction rules of the local semantics can be 
found in Figure~\ref{fig:operators}. Here,
we consider that a local state has the form
$\theta,e$, where $\theta$ is the current environment 
(a variable substitution) and $e$ is an
expression (to be evaluated). 

\begin{figure}[t]
\centering
$
  \begin{array}{rc}
    (\mathit{Check})~~~~~~ & {\displaystyle
      %\frac{} %\kappa~\mbox{is a fresh symbol}}
      {\theta,\mathsf{check}() \arro{\mathsf{check}(\kappa)}
      \theta,\kappa}
      }\\[1ex]

    (\mathit{Commit})~~~~~~ & {\displaystyle
      %\frac{}
      {\theta,\mathsf{commit}(\tau) \arro{\mathsf{commit(\tau)}}
      \theta,\mathit{ok}}
      }\\[1ex] 
      
    (\mathit{Rollback})~~~~~~ & {\displaystyle
      %\frac{}
      {\theta,\mathsf{rollback}(\tau) \arro{\mathsf{rollback}(\tau)}
      \theta,\mathit{ok}}
      } %\\[2ex]
  \end{array}
$
\caption{Rollback recovery operators} \label{fig:operators}
\end{figure}

Rule \emph{Check} reduces the call to a future, $\kappa$, which
also occurs in the label of the transition step. 
As we will see in the next section, 
the corresponding rule in the system semantics will perform
the associated side-effect (creating a checkpoint) and will also
bind $\kappa$ with the (fresh) identifier for this checkpoint.		

Rules \emph{Commit} and \emph{Rollback} pass the corresponding 
information to the system semantics in order to do the 
associated side effects. 
Both rules reduce the call to the constant ``$\mathit{ok}$'' 
(an atom commonly used in Erlang when a function call 
does not return any value).

In the following, we extend the notation in the previous
section with three new actions:
\blue{$\tau \!\leftarrow\! \mathsf{check}()$},
\blue{$\mathsf{commit}(\tau)$}, and
\blue{$\mathsf{rollback}(\tau)$}, with the obvious meaning.

\begin{figure}[t]
\begin{verbatim}
  bank(B) -> T = check(),
             receive
               {C,get} -> 
                   C ! B, commit(T), bank(B);
               {C,withdraw,N} -> 
                   try 
                      C ! ack,
                      ...  // some safety checks
                      commit(T), C ! ok, bank(NB)
                   catch
                      _:_ -> rollback(T),bank(B)
                   end
             end.
\end{verbatim}
\caption{Example~\ref{ex:one} including rollback recovery operators}
\label{fig:program2}
\end{figure}

\begin{example} \label{ex:three}
  Consider again the program in Example~\ref{ex:one}, 
  where function \verb$bank/1$ is now modified as
  shown in Figure~\ref{fig:program2}.
  Assuming that the safety checks fail and the bank account
  server calls $\mathsf{rollback}$ (instead of $\mathsf{commit}$),
  the sequence of actions for process \verb$bank$ are now
  the following:\\[1ex] 
  \blue{$
  \begin{array}{l}
  \red{\tau_1\!\leftarrow\!\mathsf{check}()};
  \mathsf{rec}({\tt get});
  \mathsf{send}({\tt client},B);
  \red{\mathsf{commit}(\tau_1)};\\
  \red{\tau_2 \!\leftarrow\! \mathsf{check}()};
  \mathsf{rec}( {\tt withdraw});
  \mathsf{send}({\tt client},{\tt ack});  
  \red{\mathsf{rollback}(\tau_2)}
  \end{array}
  $}\\[1ex]
\begin{figure}[t]
\centering
\begin{tikzpicture}
\draw[->,dashed,thick] (0,4) node[above]{\tt init} -- (0,0);
\draw[->,dashed,thick] (3,4) node[above]{\tt bank} -- (3,0);
\draw[->,dashed,thick] (6,4) node[above]{\tt client} -- (6,0);

\draw[->,dotted] (0,3.7) node[left]{$\mathsf{spawn}$} -- (3,3.6);

\draw[->,dotted] (0,3.3) node[left]{$\mathsf{spawn}$} -- (6,3.1);

\draw (3,2.9) node[left]{\red{$\mathsf{check}$}}
node[right]{(\red{$\tau_1$})};

\draw[->] (6,2.7) node[right]{$\mathsf{send}$} -- (3,2.5)
node[midway,above]{\tt get} node[left]{$\mathsf{rec}$};

\draw[->] (3,2.2) node[left]{$\mathsf{send}$} -- (6,2.0)
node[midway,above]{$B$} node[right]{$\mathsf{rec}$};

\draw (3,1.8) node[left]{\red{$\mathsf{commit}$}}
node[right]{(\red{$\tau_1$})};

\draw (3,1.4) node[left]{\red{$\mathsf{check}$}}
node[right]{(\red{$\tau_2$})};

\draw[->] (6,1.2) node[right]{$\mathsf{send}$} -- (3,1.0)
node[midway,above]{\tt withdraw} node[left]{$\mathsf{rec}$};

\draw[->] (3,0.7) node[left]{$\mathsf{send}$} -- (6,0.5)
node[midway,above]{\tt ack} node[right]{$\mathsf{rec}$};

\draw (3,0.3) node[left]{\red{$\mathsf{rollback}$}}
node[right]{(\red{$\tau_2$})};
\end{tikzpicture}

\vspace{1ex}

\mbox{}\hspace{-12ex}{\blue{{\Large $\Downarrow$} after $\mathsf{rollback}(\tau_2)$}}

\vspace{1ex}

\begin{tikzpicture}
\draw[->,dashed,thick] (0,3.2) node[above]{\tt init} -- (0,0);
\draw[->,dashed,thick] (3,3.2) node[above]{\tt bank} -- (3,0);
\draw[->,dashed,thick] (6,3.2) node[above]{\tt client} -- (6,0);

\draw[->,dotted] (0,2.9) node[left]{$\mathsf{spawn}$} -- (3,2.8);

\draw[->,dotted] (0,2.5) node[left]{$\mathsf{spawn}$} -- (6,2.3);

\draw[->] (6,1.9) node[right]{$\mathsf{send}$} -- (3,1.7)
node[midway,above]{\tt get} node[left]{$\mathsf{rec}$};

\draw[->] (3,1.4) node[left]{$\mathsf{send}$} -- (6,1.2)
node[midway,above]{$B$} node[right]{$\mathsf{rec}$};

\draw[->,dashed] (6,0.8) node[right]{$\mathsf{send}$} -- (3,0.6)
node[midway,above]{\tt withdraw} node[left]{};

\draw (3,0.3) node[left]{\red{$\mathsf{check}$}}
node[right]{(\red{$\tau_3$})};

%\draw (3,0.6) node[left]{$\mathsf{rec}$}
%node[right]{};

\end{tikzpicture}
\caption{Graphical representation of the execution in
Example~\ref{ex:three}  (time flows from top to bottom)}
% (time flows from top to bottom)}
\label{fig:program2-rollback}
\vspace{-2ex}
\end{figure}
A graphical representation of the new execution
can be found in Figure~\ref{fig:program2-rollback}. 
Intuitively speaking, it proceeds as follows:
\begin{itemize}
\item The bank account server calls 
$\mathsf{check}()$ at the start of each cycle,
creating a checkpoint with the current state of the process. 
In the first call, it returns 
$\tau_1$ (so $T$ is bound to $\tau_1$). 

\item Since the ``get'' operation completes successfully,
we have $\mathsf{commit}(\tau_1)$ which removes the saved
checkpoint and the current state becomes irreversible.
A recursive call to $\mathsf{bank}(B)$ starts a 
new cycle.

\item The next cycle starts by creating a new checkpoint
$\tau_2$. After the withdrawal request, the 
server sends an ack to the client. Here, 
we assume that something bad happens and an exception 
is raised.
Therefore, execution jumps to the rollback operation,
which recovers the state at checkpoint $\tau_2$ but
now calls $\mathsf{bank}(B)$ with the old balance.
Furthermore, all causally dependent operations are
undone too (the case of the receiving of message
{\tt ack} in the client process).

\item Finally, the call to $\mathsf{bank}(B)$ starts a
new cycle, which creates a new checkpoint ($\tau_3$), 
and so forth. Note that the client does not need to resend
the withdrawal request, since the rollback operation
will put the message back on the network.
\end{itemize}
\end{example}

%%%%%%%%%%%%%%%%%%%%%%%%%%%%%%%%%%%%%
\section{Designing an Asynchronous Rollback Recovery Strategy} \label{sec:design}

Let us now consider the design of a \emph{practical}
rollback recovery strategy. In principle, we have the
following requirements:
\begin{enumerate}
\item First, rollback recovery should be performed without
the need of a central coordination. For practical applications,
it would be virtually impossible 
to synchronize all processes, especially the remote ones.
This implies that every process interaction must be
based on (asynchronous) message-passing.

\item Secondly, recovery must bring the system to a 
\emph{consistent} global state. For this purpose, 
we will propagate checkpoints following the causal 
dependencies of a process. In particular, every process
spawning or message sending will introduce a 
\emph{forced} checkpoint 
(following the terminology of \cite{EAWJ02}). 
Consequently, if a process rolls back
to a given checkpoint, we might have to also roll back
other processes to the respective (forced) checkpoints.
\end{enumerate}
In order to materialize this strategy, we extend
the standard semantics to store the checkpoints
of each process as well as some information regarding its
actions; namely, we add a \emph{history} containing
a list of the following elements: 
\red{$\mathit{check}(\tau,s)$}, where $\tau$ is a checkpoint
identifier and $s$ is a state; \red{$\mathit{send}(p,\ell)$},
where $p$ is a pid and $\ell$ is a message tag
(see below); \red{$\mathit{rec}(\cC,p,p',\{\ell,v\})$}, where
$\cC$ is a set of checkpoint identifiers, $p,p'$
are pids,
and $\{\ell,v\}$ is a message $v$ tagged with $\ell$;
and \red{$\mathit{spawn}(p)$}, where $p$ is a pid.

\begin{definition}[extended process configuration]
  An extended process configuration is denoted by a tuple of 
  the form $\tuple{\Delta,p,s}$, where 
  $\Delta$ is a history, 
  $p$ is the pid of the process
  and $s$ is its current state.
\end{definition}
In the following, we let $\nil$ denote an empty list 
and $x\cons xs$ a list with head $x$ and tail $xs$. 
Messages are now extended in two ways. First, message values are
wrapped with a tag so that they can be uniquely
identified (as in \cite{LPV21}).
And, secondly, they now include the set
of \emph{active} checkpoints of the sender so that they
can be propagated to the receiver (as
\emph{forced} checkpoints):

\begin{definition} %[extended message]
  An \emph{extended message} 
  has the form $(\cC,p,p',\{\ell,v\})$, 
  where $\cC$ is a set of checkpoint
  identifiers, $p$ is the pid 
  of the sender, $p'$ that of the receiver, and $\{\ell,v\}$ 
  is a tagged value. 
\end{definition}
Besides ordinary (extended) 
messages, we also introduce a new kind
of messages, called system notifications:

\begin{definition}[system notification]
  A \emph{system notification} 
  has the form $\tuplefb{p,p',v}$, where 
  $p$ is the pid of the sender, 
  $p'$ that of the receiver, and $v$ is the 
  message value.\footnote{We note that system notifications
  are not tagged since they will never be undone.} 
\end{definition}
This new kind of messages is necessary since, according to
the standard semantics, delivered messages are not processed
unless there is a corresponding receive statement.
In our strategy, though, we might need to send a notification
to a process at any point in time. This is why system
notifications are needed. An implementation of this strategy
could be carried over using 
run-time monitors (as in the reversible 
choreographies of \cite{FMT18}).

In the following, a system is given by the parallel 
composition of extended process configurations, messages,
and system notifications.
Before presenting the instrumented semantics for rollback
recovery, there is one more issue which is worth discussing.
The strategy sketched above will not always work if we 
accept \emph{nested} checkpoints. 
For instance, given a sequence
of actions like\\[1ex]
\blue{$
\hspace{3ex}\ldots,\tau_1\!\leftarrow\!\mathsf{check}(),
\ldots,\tau_2\!\leftarrow\!\mathsf{check}(),
\ldots,\mathsf{commit}(\tau_1),
\ldots
$}\\[1ex]
we might have a problem if a subsequent call to
$\mathsf{rollback}(\tau_2)$ is produced. In general,
we cannot delete the saved state of a checkpoint 
($\tau_1$ above) if there is some other active
checkpoint ($\tau_2$ above) whose rollback would 
require the (deleted!) saved state. In order to overcome
this drawback, there are several possible solutions:
\begin{itemize}
\item We can forbid (or delay) checkpoints (either proper
or forced) when there is already an active checkpoint
in a process. Although some works avoid nested checkpoints
(e.g., \cite{MTY23}), we consider it overly restrictive.
\item As an alternative, we propose the \emph{delay}
of commit operations when a situation like the above one
is produced. This strategy has little impact in practice
since the checkpoint responsible for the delay will
typically either commit or roll back in a short lapse
of time.
\end{itemize}
In the following, we let \red{$\overline{\mathit{check}}(\tau,s)$}
denote a checkpoint whose commit operation 
has been \emph{delayed}
(and, thus, is not active anymore). % (but cannot be deleted yet).

%%%%%%%%%%%%%%%%%%%%%%%%%%%%%%%%%%%%%
\section{Rollback Recovery Semantics}

In this section, we introduce a labeled transition relation,
$\hoo$, that formally specifies our rollback recovery strategy.

\begin{figure}[tbh]
\centering
$\begin{array}{r@{~~}c}
    ({\mathit{Check}}) & {\displaystyle
        \frac{s \arro{\mathsf{check}(\kappa)} s'~\mbox{and}~
        {\tau~\mbox{is a fresh identifier}}
      }  
    {\tuple{\Delta,p,s} \hoo
           \tuple{\red{\mathit{check}(\tau,s)}\cons\Delta,p,
           s'[\kappa\leftarrow \tau]}}
      }
\end{array}
$
\caption{Rollback recovery semantics: $\mathsf{check}()$} 
\label{fig:rollback-semantics-check}
\end{figure}

The first rule, \emph{Check}, introduces a new checkpoint in 
the history of a process (Figure~\ref{fig:rollback-semantics-check}),
where $s$ denotes the saved state.

\begin{figure*}[t]
\centering
$\begin{array}{r@{~~}c}
      ({\mathit{Seq}}) & {\displaystyle
      \frac{s \arro{\mathsf{seq}} s'
      }{\tuple{\Delta,p,s} 
        \hoo 
        \tuple{\Delta,p,s'}}
      } 
\hspace{5ex}
    ({\mathit{Send}}) ~ {\displaystyle
      \frac{s \arro{\mathsf{send}(p',v)} s',~
      {C = \mathsf{chks}(\Delta)},~\mbox{and}~
     {\ell~\mbox{is fresh}}
      }{\tuple{\Delta,p,s} 
        \hoo
         ({\cC},p,p',\{\ell,v\}) \comp 
         \tuple{{\mathsf{add}(\red{send(p',\ell)},\Delta)},p,s'}}
      }\\[3ex]
      
      ({\mathit{Receive}})   
&  {\displaystyle
        \frac{%\begin{array}{l}
        s \arro{\mathsf{rec}(\kappa,cs)}
          s',~ \mathsf{matchrec}(cs,v) = cs_i, %%\\
          %\hspace{0ex}
          {\cC = \mathsf{chks}(\Delta)},~\mbox{and}
          ~{\cC'\setminus
          \cC=\{\tau_1,\ldots,\tau_n\}}
          %\end{array}
         }
          {%\begin{array}{l}
          ({\cC'},p',p,\{\ell,v\}) \comp
           \tuple{\Delta,p,s} \hoo  %%\\ 
           \tuple{\mathsf{add}(\red{\mathit{rec}(\red{\cC'},p',p,\{\ell,v\})},\red{\mathit{check}(\tau_1,s)}\cons\ldots\cons\red{\mathit{check}(\tau_n,s)}\cons\Delta),p,s'[\kappa\leftarrow cs_i]}
           %\end{array}
           }
      }\\[3ex]  
      
   ({\mathit{Spawn}}) & {\displaystyle
        \frac{%\begin{array}{l}
        s \arro{\mathsf{spawn}(\kappa,s_0)} 
          s',~ p'~\mbox{is a fresh pid},  %\\
          ~\mbox{and}~
          {\mathsf{chks}(\Delta)=\{\tau_1,\ldots,\tau_n\}}
          %\end{array}
          }
          {%\begin{array}{l}
          \tuple{\Delta,p,s} \hoo    
           \tuple{{\mathsf{add}(\red{spawn(p')},\Delta)},p,s'[\kappa\leftarrow p']}  %%\\
           %\hspace{5ex}
           \comp 
           \tuple{[\red{\mathit{check}(\tau_1,\bot)},\ldots,\red{\mathit{check}(\tau_n,\bot)}],p',s_0}
           %\end{array}
           }
      }\\[3ex]

      ({\mathit{Par}}) & {\displaystyle
      \frac{S_1 \hoo S'_1~~\mbox{and}~~\id(S'_1)\cap \id(S_2)=\emptyset}{S_1  \comp S_2 
         \hoo S'_1 \comp S_2 
        }
      }      

  \end{array}
  $
\caption{Rollback recovery semantics: core rules} 
\label{fig:rollback-semantics-core}
\end{figure*}

Consider now the extension of the rules in the standard
semantics to deal with checkpoints 
(Figure~\ref{fig:rollback-semantics-core}).
Rules \emph{Seq} and \emph{Par} 
are extended in a trivial way since the history
is not modified. In rule \emph{Par}, we assume now 
that $\id(S)$ returns, not only the set of pids, 
but also the set of message tags in $S$.

\begin{figure}[t]
\centering

\hspace{1ex} 
$
\mathsf{chks}(\Delta) =
\left\{\begin{array}{ll}
\emptyset & \mbox{if}~\Delta =\nil\\
\tau\cup\mathsf{chks}(\Delta') 
&\mbox{if}~\Delta=\mathit{check}(\tau,s)\cons\Delta'\\
\mathsf{chks}(\Delta') 
&\mbox{if}~\Delta=\overline{\mathit{check}}(\tau,s)\cons\Delta'\\
\end{array}
\right.
$

\vspace{2ex}

$
\begin{array}{l}
\mathsf{add}(a,\Delta) = 
\left\{\begin{array}{ll}
\Delta & \mbox{if}~\mathsf{chks}(\Delta) = \emptyset\\
a\cons\Delta & \mbox{otherwise}
\end{array}
\right.
\end{array}
$
\caption{Auxiliary functions (I)} \label{fig:auxiliary-functions}
\end{figure}

As for rule \emph{Send}, we perform several extensions.
As mentioned before, every message $v$ is now wrapped with 
a (fresh) tag $\ell$ and, 
%This is necessary to uniquely identify every message.
moreover, it includes the set of (active)
checkpoint identifiers, $\cC$, which is computed using 
function $\mathsf{chks}$ (Figure~\ref{fig:auxiliary-functions}).
Finally, we add a new element, $\mathsf{send}(p',\ell)$, 
to the history if there is at least one
active checkpoint. We use the auxiliary function $\mathsf{add}$
for this purpose (Figure~\ref{fig:auxiliary-functions}). 
%Both auxiliary functions, $\mathsf{chks}$ and
%$\mathsf{add}$, can be found in Figure~\ref{fig:auxiliary-functions}.

Rule \emph{Receive} proceeds now as follows. First, we
take the set of checkpoint identifiers from the message and
delete those which are already active in the process.
If there are no new checkpoints 
($C'\setminus C = \emptyset$ and $n=0$), this rule
is trivially equivalent to the same rule in the standard semantics.
Otherwise, a number of new \emph{forced} checkpoints
are introduced.

Finally, rule \emph{Spawn} is extended in two ways:
a new element is added to the process history (assuming there
is at least one active checkpoint) and
the spawned process is initialized with
a number of forced checkpoints, one for each active checkpoint
in process $p$. Here, $\bot$ is used as a special ``null''
value which will be useful later to detect that the process
must be deleted in case of a rollback.

It is easy to see that 
the rollback recovery semantics so far is a
conservative extension of the standard semantics:
if there are no active checkpoints, the rules in
Figure~\ref{fig:rollback-semantics-core} are equivalent
to those in Figure~\ref{fig:standard-semantics}.

\begin{figure*}[tbh] 
  \centering
  $
  \begin{array}{r@{~~}c}
      ({\mathit{Rollback}}) & {\displaystyle
      \frac{s \arro{\mathsf{rollback}(\tau)} s'
      ~\mbox{and}~
      \mathsf{chk}(\tau,\Delta) =
      (\Delta',s_\tau,L,\{p_1,\ldots,p_n\},\mathit{Ms})
      }{\tuple{\Delta,p,s} 
        \hoo
        \tuple{\Delta',p,s'\oplus s_\tau}^{\tau,p,p,L,\{p_1,\ldots,p_n\}}
        \comp
        \tuplefb{p,p_1,\{p,\mathit{roll},\tau\}}\comp\ldots\comp
        \tuplefb{p,p_n,\{p,\mathit{roll},\tau\}}\comp \mathit{Ms}
        }
      }    \\[5ex]

      ({\mathit{Roll1}}) & {\displaystyle
      \frac{\tau\in\Delta ~~\mbox{and}~~\mathsf{chk}(\tau,\Delta) =
      (\Delta',s_\tau,L,\{p_1,\ldots,p_n\},\mathit{Ms})
      }{\tuplefb{p',p,\{p_\tau,\mathit{roll},\tau\}}\comp
        \tuple{\Delta,p,s} 
        \hoo
        \tuple{\Delta',p,s_\tau}^{\tau,p_\tau,p',L,\{p_1,\ldots,p_n\}}
        \comp
        \tuplefb{p,p_1,\{p_\tau,\mathit{roll},\tau\}}\comp\ldots\comp
        \tuplefb{p,p_n,\{p_\tau,\mathit{roll},\tau\}}\comp \mathit{Ms}
        }
      }    \\[3ex]

      ({\mathit{Roll2}}) & {\displaystyle
      \frac{\tau\not\in\Delta
      }{\tuplefb{p',p,\{p_\tau,\mathit{roll},\tau\}}\comp
        \tuple{\Delta,p,s} 
        \hoo
        \tuple{\Delta,p,s}^{\tau,p_\tau,p',\emptyset,\emptyset}
        %\comp
        %\tuplefb{p,p',\{\mathit{done\mbox{-}async},\tau\}}
        }
      }    \\[3ex]

      ({\mathit{Roll3}}) & {\displaystyle
      \tuplefb{p',p,\{p_\tau,\mathit{roll},\tau\}}
      \comp\tuple{\Delta,p,s}^{\tau',p_1,p_2,L,P}
      \hoo
      \tuple{\Delta,p,s}^{\tau',p_1,p_2,L,P}
      \comp \tuplefb{p,p',\{\mathit{done\mbox{-}async},\tau\}}\hspace{5ex}\mbox{if}~\tau'\leqslant_\Delta \tau
      }    \\[3ex]

      ({\mathit{Undo\mbox{-}send}}) & {\displaystyle
      (\cC,p',p,\{\ell,v\})\comp
      \tuple{\Delta,p,s}^{\tau,p_\tau,p_r,L,P}
      \hoo
      \tuple{\Delta,p,s}^{\tau,p_\tau,p_r,L\setminus\{\ell\},P}\hspace{5ex}\mbox{if}~\ell\in L
      }    \\[1ex]

      ({\mathit{Undo\mbox{-}dep1}}) & {\displaystyle
      \tuplefb{p',p,\{\mathit{done\mbox{-}async},{\tau}\}}\comp
      \tuple{\Delta,p,s}^{{\tau},p_\tau,p_r,\emptyset,P}
      \hoo
      \tuple{\Delta,p,s}^{\tau,p_\tau,p_r,\emptyset,P\setminus\{p\}}
      %\comp \tuplefb{p',p,\{\mathit{resume},\tau\}}
      }    \\[1ex]

      ({\mathit{Undo\mbox{-}dep2}}) & {\displaystyle
      \tuplefb{p',p,\{\mathit{done\mbox{-}sync},{\tau}\}}\comp
      \tuple{\Delta,p,s}^{{\tau},p_\tau,p_r,\emptyset,P}
      \hoo
      \tuple{\Delta,p,s}^{\tau,p_\tau,p_r,\emptyset,P\setminus\{p\}}
      \comp \tuplefb{p,p',\{\mathit{resume},\tau\}}
      }    \\[3ex]

      ({\mathit{Resume1}}) & {\displaystyle
      \tuple{\Delta,p,s}^{\tau,p_\tau,p_r,\emptyset,\emptyset}
      \hoo
      \tuple{\Delta,p,s}\hspace{5ex}\mbox{if}~p= p_\tau
      }    \\[1ex]
      
      ({\mathit{Resume2}}) & {\displaystyle
      \tuple{\Delta,p,\bot}^{\tau,p_\tau,p_r,\emptyset,\emptyset}
      \hoo
      \tuplefb{p,p_r,\{\mathit{done\mbox{-}async},\tau\}}\hspace{5ex}\mbox{if}~p\neq p_\tau
      }    \\[1ex]

      ({\mathit{Resume3}}) & {\displaystyle
      \tuple{\Delta,p,s}^{\tau,p_\tau,p_r,\emptyset,\emptyset}
      \hoo
      \tuple{\Delta,p,s}^{\tau}\comp
      \tuplefb{p,p_r,\{\mathit{done\mbox{-}sync},\tau\}}\hspace{5ex}\mbox{if}~p\neq p_\tau
      ~\mbox{and}~s\neq \bot
      }    \\[1ex]
      
      ({\mathit{Resume4}}) & {\displaystyle
      \tuplefb{p',p,\{\mathit{resume},\tau\}}\comp
      \tuple{\Delta,p,s}^{\tau}
      \hoo \tuple{\Delta,p,s}
      %\hspace{5ex}\mbox{if}~p\neq p'
      }    \\
  \end{array}
  $
\caption{Rollback recovery semantics: rollback rules} 
\label{fig:rollback-semantics-rollback}
\end{figure*}

\begin{figure*}[thb]
$
\mathsf{chk}(\tau,\Delta) =
\left\{\begin{array}{ll}
(\Delta',s,\emptyset,\emptyset,\emptyset) & \mbox{if}~\Delta = \mathit{check}(\tau,s)\cons\Delta'\\
(\Delta',s,L,P\cup\{p\},\mathit{Ms}) & \mbox{if}~\Delta = \mathit{spawn}(p)\cons\Delta'~\mbox{and}~ 
\mathsf{chk}(\tau,\Delta') = (\Delta',s,L,P,\mathit{Ms})\\
(\Delta',s,L\cup\{\ell\},P\cup\{p\},\mathit{Ms}) & \mbox{if}~\Delta = \mathit{send}(p,\ell)\cons\Delta'~\mbox{and}~ 
\mathsf{chk}(\tau,\Delta') = (\Delta',s,L,P,\mathit{Ms})\\
(\Delta',s,L,P,\mathit{Ms}\cup\{(\cC,p,p',\{\ell,v\})\}) 
& \mbox{if}~\Delta = \mathit{rec}(\cC,p,p',\{\ell,v\})\cons\Delta'~\mbox{and}~ 
\mathsf{chk}(\tau,\Delta') = (\Delta',s,L,P,\mathit{Ms})\\
\mathsf{chk}(\tau,\Delta') 
& \mbox{otherwise, with}~\Delta = a\cons \Delta'\\
\end{array}\right.
$\\[1ex]
\begin{minipage}{.45\linewidth}
$
\mathsf{last}(\tau,\Delta) = \left\{\begin{array}{ll}
\mathit{true} & \mbox{if}~\Delta = \mathit{check}(\tau,s)\cons\Delta'\\
\mathit{false} & \mbox{if}~\Delta = \mathit{check}(\tau',s)\cons\Delta',~\tau\neq\tau'\\
\mathsf{last}(\tau,\Delta') &\mbox{otherwise, with}~\Delta=a\cons\Delta'
\end{array}
\right. 
$\\[1ex]
$
\mathsf{dp}(\tau,\Delta) = P~~\mbox{if}~~
\mathsf{chk}(\tau,\Delta) = (\Delta',s,L,P,\mathit{Ms})
$\\[1ex]
$
\mathsf{del}(\tau,\Delta) = \left\{\begin{array}{ll}
\Delta' & \mbox{if}~\Delta = \mathit{check}(\tau,s)\cons\Delta'\\
\mathsf{del}(\tau,\Delta') &\mbox{otherwise, with}~\Delta=a\cons\Delta'
\end{array}
\right. 
$
\end{minipage}
~~~
\begin{minipage}{.45\linewidth}
$
\mathsf{delay}(\tau,\Delta) = \left\{\begin{array}{ll}
\overline{\mathit{check}}(\tau,s)\cons\Delta' & \mbox{if}~\Delta = \mathit{check}(\tau,s)\cons\Delta'\\
a\cons\mathsf{delay}(\tau,\Delta') &\mbox{otherwise, with}~\Delta=a\cons\Delta'
\end{array}
\right. 
$\\[1ex]
$
\mathsf{delayed}(\Delta) = \left\{\begin{array}{ll}
\emptyset & \mbox{if}~\Delta = \nil\\
%\emptyset & \mbox{if}~\Delta = \mathit{check}(\tau,s)\cons\Delta' \\
\{\tau\} & \mbox{if}~\Delta = \overline{\mathit{check}}(\tau,s)\cons\Delta' \\
\mathsf{delayed}(\Delta') &\mbox{otherwise, with}~\Delta=a\cons\Delta'
\end{array}
\right. 
$
\end{minipage}
\caption{Auxiliary functions (II)} \label{fig:auxiliary-functions-2}
\end{figure*}

Let us now consider the rollback rules 
(Figure~\ref{fig:rollback-semantics-rollback}).
Roughly speaking, 
the execution of a rollback involves the following
steps:
\begin{itemize}
\item First, the process is \emph{blocked} so that the 
forward rules (Figures~\ref{fig:rollback-semantics-check} 
and \ref{fig:rollback-semantics-core}) cannot be applied.
In particular, when a process configuration is adorned with
some superscripts, the forward rules are not applicable.
\item Then, the process recovers the state saved in the 
checkpoint and puts all received messages (since the checkpoint
occurred) back on the network. 
\item The rollback is then propagated to all 
processes where a forced checkpoint might have been introduced:
spawned processes and recipients of a message 
(since the checkpoint). % occurred).
\item Finally, the process keeps waiting for the 
rollbacks of these
forced checkpoints to complete in order to resume its
normal, forward computation.
\end{itemize}
This process is formalized in rule $\mathit{Rollback}$,
where function $\mathsf{chk}$ takes a checkpoint identifier
$\tau$ and a history $\Delta = a_1\cons \ldots\cons a_m\cons
\mathit{check}(\tau,s_\tau)\cons\Delta'$ and returns a 
tuple $(\Delta',s_\tau,
L,P,\mathit{Ms})$, where $L$ is a set with the tags of the 
sent messages,
%(i.e., $\ell\in L$ if there is some 
%$a_i=\mathit{send}(p',\ell)$, $i\in\{1,\ldots,m\}$), 
$P$ includes the pids of the spawned processes as well as
the pids of the message recipients, 
%(i.e., $p\in P$ if there is some
%$a_i = \mathit{spawn}(p)$ or $\mathit{send}(p,\ell)$,
%$i\in\{1,\ldots,m\}$), 
and $\mathit{Ms}$ is a set with the
received messages (all of them within $a_1,\ldots,a_m$).
%(i.e., $(\cC,p,p',\{\ell,v\})\in \mathit{Ms}$ 
%if there is
%some $a_i = \mathit{rec}(\cC,p,p',\{\ell,v\})$, 
%$i\in\{1,\ldots,m\}$). 
The function definition can be found in
Figure~\ref{fig:auxiliary-functions-2}.
The blocked configuration is adorned with the
superscripts $\tau,p,p,L,P$. The pid $p$ is duplicated
since we will later distinguish the pid of the process
that started the rollback from that of the 
process propagating it (which may be different, see below).
Moreover, note that rule $\mathit{Rollback}$ does not 
recover the saved state $s_\tau$
but $s'\oplus s_\tau$. Nevertheless, in this paper,
we assume that $s'\oplus s_\tau = s_\tau$ for simplicity.
However, other definitions of  ``$\oplus$'' are
possible. For instance, we might
have $s'\oplus s_\tau = (\theta_\tau,e')$ if we want to
combine the saved environment with the next expression
to be evaluated (as in Example~\ref{ex:three}), where
$s'=(\theta',e')$ and $s_\tau = (\theta_\tau,e_\tau)$.

Rollback requests are propagated to other processes by means
of system notifications of the following form: 
$\tuplefb{p,p_i,\{p_\tau,roll,\tau\}}$. This is dealt with rules
$\mathit{Roll1}$, $\mathit{Roll2}$, and $\mathit{Roll3}$. 
If the system notification reaches a process in normal,
forward mode and the (forced) checkpoint exists (denoted
with $\tau\in\Delta$), then rule $\mathit{Roll1}$ proceeds almost
analogously to rule $\mathit{Rollback}$; the main difference
is that 
%the recovered state is the one stored in the 
%forced checkpoint (no alternative here), and that
the superscripts with the pids of the process that started
the rollback and the one that sent the system notification
are generally different.
Rule $\mathit{Roll2}$ applies when
the checkpoint does not exist ($\tau\not\in\Delta$), e.g.,
because the message propagating the rollback was not
yet received. In this case, we still block the process
to avoid it receives the message before the rollback
is complete. 
%but send immediately a system notification of the form
%$\tuplefb{p,p',\{\mathit{done\mbox{-}async},\tau\}}$ back to 
%process $p'$ (the one that started the rollback).
%
There are two kinds of system notifications to let
a process know that the requested rollback has been
completed: $\mathit{done\mbox{-}async}$ is asynchronous
and expects no reply, while 
$done\mbox{-}sync$ keeps the process waiting for a
reply before resuming its computation.
Finally, if the process is already in rollback
mode, we distinguish two cases: if the ongoing
rollback is older, denoted by $\tau'\leqslant_\Delta \tau$,
the rollback is
considered ``done'' and rule $\mathit{Roll3}$ sends a 
system notification of the form
$\tuplefb{p,p',\{\mathit{done\mbox{-}async},\tau\}}$ back to 
process $p'$;
otherwise (i.e., if the requested rollback is older than the
ongoing one), the rollback request is ignored until the process
ends the current 
rollback.\footnote{Note that a deadlock is not possible,
no matter if we have processes with mutual dependencies. 
Consider, e.g., two processes, $p_1$ and $p_2$, such that
each process $p_i$ creates a checkpoint $\tau_i$, sends
a message tagged with $\ell_i$ addressed to the other process
and, finally, starts a rollback 
to checkpoint $\tau_i$. In this case,
it might be the case that message $\ell_1$ reaches $p_2$
before checkpoint $\tau_2$ or $\ell_2$ reaches $p_1$
before checkpoint $\tau_1$, but both things are not possible 
at the same time (a message would need
to travel back in time).}

In order for a blocked process to resume its forward computation,
all sent messages ($L$) must be deleted from the network, and
all process dependencies ($P$) corresponding to forced
checkpoints must be completed. This is dealt with rules
$\mathit{Undo}\mbox{-}\mathit{send}$,
$\mathit{Undo}\mbox{-}\mathit{dep1}$,
and $\mathit{Undo}\mbox{-}\mathit{dep2}$. 
Note that rule $\mathit{Undo}\mbox{-}\mathit{dep2}$ sends
a system notification of the form 
$\tuplefb{p,p',\{\mathit{resume},\tau\}}$
back to process $p'$ so that it can
resume its normal computation (in contrast to rule
$\mathit{Undo}\mbox{-}\mathit{dep1}$). 
This additional communication is needed to avoid 
a situation where a process resumes its execution 
and receives again the messages that were put back 
on the network before they can be deleted 
by the process propagating the rollback.

Finally, once both $L$ and $P$ are empty (i.e., all sent
messages are undone and the rollbacks of all associated 
forced checkpoints are completed), we can apply the 
$\mathit{Resume}$ rules. Rule $\mathit{Resume1}$ applies when
the process is the one that started the rollback, and it
simply removes the superscripts.
Rule $\mathit{Resume2}$ applies when the forced
checkpoint was introduced by a process spawning and,
%the process was spawned
%after the checkpoint $\tau$ and, 
thus, it is deleted from the
system (and a system notification is sent back to $p$).
Otherwise (the case of
a process with a forced checkpoint associated to
a message receiving), we proceed in two steps: first, rule
$\mathit{Resume3}$ sends a system notification to process $p_r$
(the one that propagated the rollback) but remains blocked;
then, once it receives a system notification of the form
$\tuplefb{p_r,p,\{\mathit{resume},\tau\}}$, rule 
$\mathit{Resume4}$ resumes its normal, forward computation.

Roughly speaking, a user-defined rollback 
will follow a sequence of the form
\blue{\[
\mathit{Rollback}+\mathit{Undo\mbox{-}send}^*
+[\mathit{Undo\mbox{-}dep1}|\mathit{Undo\mbox{-}dep2}]^*
+\mathit{Resume1}
\]}
while a propagated rollback will
typically have the form
\blue{\[
\mathit{Roll1}+\mathit{Undo\mbox{-}send}^*
+[\mathit{Undo\mbox{-}dep1}|\mathit{Undo\mbox{-}dep2}]^*
+\mathit{Resume2}
\] }
(if the forced checkpoint is associated
to process spawning) or
\blue{\[
\mathit{Roll1}+\mathit{Undo\mbox{-}send}^*
+[\mathit{Undo\mbox{-}dep1}|\mathit{Undo\mbox{-}dep2}]^*
+\mathit{Resume3}+\mathit{Resume4}
\]}
(if the forced checkpoint is associated
to message sending). 
We also have a couple of corner cases when the process 
does not include the requested checkpoint 
(e.g., because the message was not yet received),
where we apply rules 
\blue{$\mathit{Roll2}+\mathit{Resume3}+\mathit{Resume4}$}, 
and when the 
process is already doing a \emph{larger} rollback 
(i.e., to a checkpoint which is older than the one
requested), where rule \blue{$\mathit{Roll3}$} is applied. 
Note that we are only showing the rules
applied to the main process.

\begin{figure}[tbh]
\centering
$
  \begin{array}{r@{~~}c}
  ({\mathit{Commit}}) & {\displaystyle
      \frac{\begin{array}{l}
      s \arro{\mathsf{commit}(\tau)} s',~ 
      \mathsf{last}(\tau,\Delta)=\mathit{true},~\\
      \hspace{3ex}\mbox{and}~
      \mathsf{dp}(\tau,\Delta)=\{p_1,\ldots,p_n\}
	  \end{array}}
      {\begin{array}{l}
      \tuple{\Delta,p,s} 
         \hoo 
        \tuple{\mathsf{del}(\tau,\Delta),p,s'}\comp\\
        \hspace{2ex}\tuplefb{p,p_1,\{commit,\tau\}}\comp\ldots\comp\tuplefb{p,p_n,\{commit,\tau\}}
        \end{array}}
      }\\[6ex]
      
  ({\mathit{Delay}})  & {\displaystyle
      \frac{s \arro{\mathsf{commit}(\tau)} s'~\mbox{and}~ 
      \mathsf{last}(\tau,\Delta)=\mathit{false}
	  }
      {\tuple{\Delta,p,s} 
         \hoo 
        \tuple{\mathsf{delay}(\tau,\Delta),p,s'}
        }
      }\\[3ex]
      
  ({\mathit{Commit2}}) & {\displaystyle
      \frac{
      \mathsf{last}(\tau,\Delta)=\mathit{true}~
      \mbox{and}~
      \mathsf{dp}(\tau,\Delta)=\{p_1,\ldots,p_n\}
	  }
      {\begin{array}{l}
      \tuplefb{p',p,\{\mathsf{commit,\tau}\}}\comp\tuple{\Delta,p,s} 
         \hoo 
        \tuple{\mathsf{del}(\tau,\Delta),p,s}\comp\\
        \hspace{1ex}\tuplefb{p,p_1,\{commit,\tau\}}\comp\ldots\comp\tuplefb{p,p_n,\{commit,\tau\}}
        \end{array}}
      }
      \\[6ex]
      
   ({\mathit{Delay2}})  & {\displaystyle
      \frac{\mathsf{last}(\tau,\Delta)=\mathit{false}
	  }
      {\tuplefb{p',p,\{\mathsf{commit,\tau}\}}\comp\tuple{\Delta,p,s} 
         \hoo 
        \tuple{\mathsf{delay}(\tau,\Delta),p,s}
        }
      }  \\[3ex]
      
     ({\mathit{Commit3}}) & {\displaystyle
      \frac{\begin{array}{l}
      \tau\in\mathit{delayed}(\Delta),~~
      \mathsf{last}(\tau)=\mathit{true},\\
      \hspace{3ex}\mbox{and}~
      \mathsf{dp}(\tau,\Delta)=\{p_1,\ldots,p_n\}
	  \end{array}}
      {\begin{array}{l}
      \tuple{\Delta,p,s} 
         \hoo 
        \tuple{\mathsf{del}(\tau,\Delta),p,s}\comp\\
        \hspace{1ex}\tuplefb{p,p_1,\{commit,\tau\}}\comp\ldots\comp
        \tuplefb{p,p_n,\{commit,\tau\}}
        \end{array}}
      }    \\
\end{array}
  $
\caption{Rollback recovery semantics: commit rules} 
\label{fig:rollback-semantics-commit}
\end{figure}

Let us finally consider the rules for $\mathsf{commit}$ 
(Figure~\ref{fig:rollback-semantics-commit}).
Basically, we have a distinction on whether the checkpoint
is the last active one or not 
(as discussed in Section~\ref{sec:design}). In the first case,
rule $\mathit{Commit}$ deletes every element in the history up
to the given checkpoint and propagates the commit operation to
all its dependencies. In the latter case, the commit operation
is delayed. Here, we use the auxiliary functions
$\mathsf{last}$, $\mathsf{del}$, and $\mathsf{delay}$, which
are defined in Figure~\ref{fig:auxiliary-functions-2}.

Rules $\mathit{Commit2}$ and $\mathit{Delay2}$ are perfectly
analogous, but the process starts by receiving a system notification
rather than a user operation.
Finally, rule $\mathit{Commit3}$ checks whether there is
some delayed commit that can be already done. This rule only
needs to be considered whenever a checkpoint is removed from
a process.

\section{Soundness of Rollback Recovery Strategy}

In this section, we consider the soundness of our approach
to rollback recovery. Essentially, we want to prove that
every step under the rollback recovery semantics is either
equivalent to an step under the standard semantics or
to an step under a causally consistent reversible semantics
for the language. Therefore, we will be able to conclude that
our rollback recovery semantics is also causally consistent
in the sense that no action is undone until all its
consequences have been already undone.

%As for the soundness of our rollback recovery strategy,
%we have proved that every derivation
%with our rollback recovery semantics can be projected to
%a causally consistent derivation under an \emph{uncontrolled} 
%reversible semantics for the language, like that in 
%\cite{LM20} or \cite{LNPV18jlamp}. See the companion
%technical report \cite{Vid24tr} for the technical details.

First, let us present the reversible semantics of
\cite{LPV19,LPV21}, where we omit the \emph{replay} component 
(using an execution trace) since it is not relevant in our context.
Furthermore, there are a few, minor differences:
\begin{itemize}
\item First, we consider floating messages and rule \emph{Par} to
lift reductions to a larger system (as in 
\cite{LSZ19,LM20}). 
%instead of using a global mailbox and
%consider the complete system in each reduction rule (as in \%cite{LNPV18jlamp,LPV21}). 
The resulting semantics is
straightforwardly equivalent but the formulation of the 
rules is simpler.

\item We consider a generic state, $s$, rather than a pair
$\theta,e$ (environment, expression) as in  \cite{LPV19,LPV21}.
This is a simple generalization to improve readability 
but does not affect the behavior of the system rules.

\item Finally, we do not consider a rule for the predefined function
\emph{self()} (that returns the pid of the current process) since it
is not relevant in the context this work (but could be added easily).
\end{itemize}

\begin{figure}[t]
  \[
  \hspace{-3ex}
  \begin{array}{r@{~~}c}
      (\underline{\mathit{Seq}}) & {\displaystyle
      \frac{s \arro{\mathsf{seq}} s'
      }{\tuple{h,p,s} 
        \rh %_{p,\underline{\mathsf{seq}}} 
        \tuple{\mathsf{seq}(s)\cons h,p,s'}}
      }\\[3ex]

    (\underline{\mathit{Send}}) & {\displaystyle
      \frac{s \arro{\mathsf{send}(p',v)} s' ~\mbox{and}~ 
      ~{\ell~\mbox{is a fresh symbol}}
      }{\tuple{h,p,s} 
        \rh %_{p,\underline{\mathsf{send}}({\ell})}
         (p,p',\{{\ell},v\}) \comp 
         \tuple{{\mathsf{send}(s,p',\ell)\cons h},p,s'}}
      }\\[3ex]
      
      (\underline{\mathit{Receive}}) &  {\displaystyle
        \frac{s \arro{\mathsf{rec}(\kappa,cs)}
          s' ~\mbox{and}~ \mathsf{matchrec}(cs,v) = cs_i
         }
          {(p',p,\{{\ell},v\}) \comp
           \tuple{h,p,s} 
           \rh %_{p,\underline{\mathsf{rec}}({\ell})}
           \tuple{{\mathsf{rec}(s,p,p',\{{\ell},v\})\cons h},p,s'[\kappa\leftarrow cs_i]}}
      }\\[3ex]  
      
   (\underline{\mathit{Spawn}}) & {\displaystyle
        \frac{s \arro{\mathsf{spawn}(\kappa,s_0)} 
          s'~\mbox{and}~ p'~\mbox{is a fresh pid}
          }
          {\tuple{h,p,s} \rh %_{p,\underline{\mathsf{spawn}}(p')}    
           \tuple{{\mathsf{spawn}(s,p')\cons h},p,s'[\kappa\leftarrow p']}\comp \tuple{\nil,p',s_0}}
      }\\[3ex]
      
    (\underline{\mathit{Check}}) & {\displaystyle
        \frac{s \arro{\mathsf{check}(\kappa)} s'~\mbox{and}~
        {\tau~\mbox{is a fresh identifier}}
      }  
    {\tuple{\Delta,p,s} \rh
           \tuple{{\mathsf{check}(\tau,s)}\cons\Delta,p,
           s'[\kappa\leftarrow \tau]}}
      }\\[3ex]

      (\underline{\mathit{Par}}) & {\displaystyle
      \frac{S_1 \rh_{l} S'_1~~\mbox{and}~~\ell(S'_1)\cap \ell(S_2)=\emptyset}{S_1  \comp S_2 
         \rh_{{l}} S'_1 \comp S_2 
        }
      }      

  \end{array}
  \]
\caption{Reversible semantics: forward rules} \label{fig:reversible-semantics-forward}
\end{figure}

\begin{figure}[t]
  \[
  \hspace{-3ex}
  \begin{array}{r@{~~}c}
    (\overline{\mathit{Seq}}) & {\displaystyle
      \tuple{\mathsf{seq}({s})\cons h,p,s'}
      \lh %_{p,\overline{\mathsf{seq}}} 
      \tuple{h,p,{s}}
      }
      \\[2ex]
  
    (\overline{\mathit{Send}}) & {\displaystyle
      (p,p',\{{\ell},v\}) \comp 
         \tuple{\mathsf{send}({s},p',{\ell})\cons h,
         p,s'} \lh %_{p,\overline{\mathsf{send}}({\ell})}         
         \tuple{h,p,{s}}
      }\\[2ex]
        
      (\overline{\mathit{Receive}}) &  {\displaystyle
      \tuple{{\mathsf{rec}({s},p,p',\{{\ell},v\})\cons h},p,s'}
      \lh %_{p,\overline{\mathsf{rec}}({\ell})}      
        (p',p,\{{\ell},v\}) \comp
           \tuple{h,p,{s}}
      }\\[2ex]
      
   (\overline{\mathit{Spawn}}) & {\displaystyle
           \tuple{\mathsf{spawn}({s},{p'})\cons h,
           p,s'}\comp \tuple{\nil,{p'},s_0} 
           \lh %_{p,\overline{\mathsf{spawn}}(p')}
       \tuple{h,p,{s}}
      }\\[2ex]

    (\overline{\mathit{Check}}) & {\displaystyle
      \tuple{\mathsf{check}(\tau,{s})\cons h,p,s'}
      \lh %_{p,\overline{\mathsf{seq}}} 
      \tuple{h,p,{s}}
      }
      \\[1ex]
      
          (\overline{\mathit{Par}}) & {\displaystyle
      \frac{S_1 \lh_{l} S'_1}{S_1  \comp S_2 
         \lh_{{l}} S'_1 \comp S_2 
        }
      }

  \end{array}
  \]
\caption{Reversible semantics: backward rules} \label{fig:reversible-semantics-backward}
\end{figure}

\noindent
The \emph{uncontrolled} reversible semantics is defined in
Figures~\ref{fig:reversible-semantics-forward} and 
\ref{fig:reversible-semantics-backward}. 
The forward rules (Fig.~\ref{fig:reversible-semantics-forward})
are similar to the forward rules of the rollback recovery
semantics in Fig.~\ref{fig:rollback-semantics-core}. Here, the
main difference is that history items store the state \emph{at
every step}. Moreover, each step adds an item to the current
history, in contrast to our rollback recovery semantics which
only does so when there is some active checkpoint.
The backward rules, however, are rather different to our
rollback rules in Figure~\ref{fig:rollback-semantics-rollback}.
On the one hand, they are \emph{uncontrolled}, which means
that they are not driven by a particular rollback request.
In other words, every causally consistent
backward execution can be proved. Furthermore, they are not always
asynchronous (e.g., the case of rule 
$\overline{\mathit{Spawn}}$,
which requires a synchronization between the process performing
the spawn and the spawned process).

The uncontrolled reversible semantics, $\rlh$, can then 
be defined as the union
of the two transition relations defined in
Figures~\ref{fig:reversible-semantics-forward} and 
\ref{fig:reversible-semantics-backward}, i.e.,
$(\rlh) \: = \: (\rh \cup \lh)$.
The causal consistency of $\rlh$ can be proved analogously
to \cite[Theorem 4.17]{LPV21} since, as mentioned above, 
there are only some minor differences.

In the following, we omit the steps with rule 
$\mathit{Par}$ (since they are the same in both semantics)
and only show reductions on the selected 
process. Furthermore, we assume a \emph{fair} selection 
strategy for processes, so that each process is eventually 
reduced. 

Now, we prove that the rollback recovery semantics is
indeed a conservative extension of the standard semantics.
For this purpose, we introduce the following 
auxiliary function $sta$ that defines a projection from 
rollback recovery configurations to standard
configurations:
%Function $sta$ takes
%a system of the rollback semantics and returns a corresponding 
%system of the standard semantics:
\[
sta(S) = \left\{ \begin{array}{ll}
	sta(S_1)\comp sta(S_2) & \mbox{if}~ S = S_1\comp S_2 \\
	\tuple{p,s} & \mbox{if}~S=\tuple{\Delta,p,s} \\
	(p,p',v) & \mbox{if}~S=(\cC,p,p',\{\ell,v\}) \\
\end{array}
\right.
\]
%where $\overline{s}$ replaces from $s$ all occurrences of
%the rollback operators ($\mathsf{check}$, $\mathsf{commit}$,
%and $\mathsf{rollback}$) and checkpoint identifiers (if any)
%by an arbitrary constant ``ok'' (an atom, using Erlang 
%terminology).
%
Functions $\mathit{sta}$ is extended to 
derivations in the obvious way. 

In the following, we consider some minimal requirements
on derivations under the rollback recovery semantics 
in order to be well-defined:
\begin{itemize}
\item As usual, the derivation should start with a 
\emph{reachable} system, where a system $S$ is
reachable if 
there exists an initial system $S_0$
of the form $\tuple{\nil,p,s}$ such that
$S_0\hoo^\ast S$.

\item The calls $\mathsf{commit}(\tau)$ and 
$\mathsf{rollback}(\tau)$ can only be made by the same
process that created the checkpoint $\tau$ (i.e., that
called $\mathsf{check}$ and was reduced to $\tau$).

\item Every call to either $\mathsf{commit}(\tau)$ or
$\mathsf{rollback}(\tau)$ must be preceded by a call
to $\mathsf{check}$ returning $\tau$.

\item A process can only have one action 
for every checkpoint $\tau$, either $\mathsf{commit}(\tau)$
or $\mathsf{rollback}(\tau)$, but not both.
%This condition can easily be verified from the history
%of a process. Note that, if a rollback (associated to a different 
%checkpoint) undoes a commit,
%it is removed from the history of the process and, thus, 
%a new commit/rollback would be possible. 
%For instance, at the end of the derivation shown in
%Example~\ref{ex3}, a new call $\mathsf{commit}(\tau_1)$
%or $\mathsf{rollback}(\tau_1)$ would be possible, since
%the initial call $\mathsf{commit}(\tau_1)$ has been undone
%by the call $\mathsf{rollback}(\tau_2)$.
\end{itemize}
%
%More formally,
%
%\begin{definition}[well-defined derivation] \label{def:welldefined}
%	A derivation $d = (S_0 \hoo S_1 \hoo \ldots \hoo S_n)$ 
%	under the rollback semantics 
%	is well-defined if the following conditions hold:
%	\begin{enumerate}
%	\item $S_0$ is a reachable configuration;
%	\item every reduction step $S_i \hoo_{p,\underline{\mathsf{commit}}(\tau)} S_{i+1}$ 
%	in $d$ is preceded by a reduction step $S_j \hoo_{p,\overline{\mathsf{check}}(\tau)} S_{j+1}$, $j<i$;
%	\item every reduction step $S_i \hoo_{p,\underline{\mathsf{rollback}}(\tau)} S_{i+1}$ 
%	in $d$ is preceded by a reduction step $S_j \hoo_{p,\overline{\mathsf{check}}(\tau)} S_{j+1}$, $j<i$;
%	\item no process history in $d$ may have both
%	$\mathsf{commit}(\tau)$ and $\mathsf{rollback}(\tau)$
%	at the same time.
%	\end{enumerate}
%\end{definition}
%%
%For instance, it is easy to prove that derivations are well-defined
%when the new operators are used to improve try\_catch as 
%proposed in the introduction\\
%
%$
%\mathtt{try}~\red{T = \mathsf{check}},~\red{X=\;} e,~\red{\mathsf{commit}(T)},~\red{X}~\mathtt{catch}~\_:\_ \rightarrow \red{\mathsf{rollback}(T)},~e'~\mathtt{end}
%$\\
%
We assume that these requirements hold
for all derivations. In particular, it would be easy to
prove that they hold when the new operators are used to
improve try\_catch as proposed before. 
In general, one could introduce a \emph{compliance check} 
at the type level to prevent undesired situations
(as in \cite{MTY23}).

%\begin{definition}[reachable system]
%	Let $S$ be a system. We say that $S$ is a reachable
%	system % (under the rollback recovery semantics) 
%	if there exists a derivation of the form
%	$\tuple{\nil,p,s} \hoo^\ast S$ 
%	under the rollback recovery semantics, 
%	for some pid $p$ and state $s$.
%\end{definition}
%
The next result states that our rollback recovery
semantics is a conservative extension of the 
standard semantics:

\begin{theorem} \label{th:sound1}
	Let $d$ be a well-defined derivation under the rollback 
	recovery semantics where only the forward rules 
	in Figure~\ref{fig:rollback-semantics-core} are applied. 
	Then, $sta(d)$ is a derivation under the
	standard semantics.
\end{theorem}

\begin{proof}
  The claim follows trivially since the components $p$ and $s$
  are the same in both semantics and, moreover, the addition
  of a history $\Delta$ in the rules of 
  Figure~\ref{fig:rollback-semantics-core} imposes no 
  additional restriction.
	\qed
\end{proof}
On top of the uncontrolled semantics, \cite{LNPV18jlamp,LPV21}
introduce a \emph{controlled} semantics that can be used
to \emph{drive} backward computations in order to satisfy
different requests, e.g.,
\begin{itemize}
\item go backwards one (or more) steps;
\item go backwards up to the introduction of a checkpoint;
\item go backwards up to the sending of a given message;
\item go backwards up to the spawning of a given process;
\item etc.
\end{itemize}
Notably, all these requests are carried over in a causal
consistent way, thus they often require other processes to
go backwards too. The rules of the controlled backward
semantics of \cite[Figure~15]{LNPV18jlamp} 
(adapted to the notation in this paper) can be
found in Figure~\ref{fig:controlled-semantics}.
Here, we only consider three kinds of rollback requests:
\begin{itemize}
\item $\tau$, which starts a (causally consistent) 
rollback until checkpoint $\tau$;
\item $\ell$, which undoes all actions up to the receiving of 
a message tagged with $\ell$; and
\item $\mathsf{sp}$, which undoes all the actions of a 
process (and then deletes it).
\end{itemize}
\begin{figure}[t]
\[
\begin{array}{lrcl}
(\mathit{Par}) & \multicolumn{3}{c}
{\displaystyle \frac{S_1 \looparrowright S'_1}
{S_1  \comp S_2 \looparrowright S'_1 \comp S_2}}\\[3ex] 

%(\mathit{Resume}) &\ttuple{h,p,s}_\emptyset
%& \leadsto & \tuple{h,p,s}\\[2ex]
%
(\mathit{Seq}) & \ttuple{\mathsf{seq}(s')\cons h,p,s}_\varphi
& \looparrowright & \ttuple{h,p,s'}_\varphi\\[2ex]

(\mathit{Check}) & \ttuple{\mathsf{check}(\tau,s')\cons h,p,s}_{\varphi}
& \looparrowright & \ttuple{h,p,s'}_{\varphi\setminus\{\tau\}}\\[2ex]

(\mathit{SP}) & \ttuple{\nil,p,s}_{\varphi\cup\{\mathsf{sp}\}} & \looparrowright & \ttuple{\nil,p,s}_\varphi\\[2ex]

(\mathit{Receive}) & \ttuple{\mathsf{rec}(s',p',p,\{\ell,v\})\cons h,p,s}_{\varphi} 
& \looparrowright & \ttuple{h,p,s'}_{\varphi\setminus\{\ell\}} 
\comp (p',p,\{\ell,v\})\\[2ex]

(\mathit{Spawn1}) & \ttuple{\mathsf{spawn}(s'',p')\cons h,p,s}_{\varphi} 
\comp \tuple{\nil,p',s'}
& \looparrowright & \ttuple{h,p,s''}_{\varphi} \\
(\mathit{Spawn2}) & \ttuple{\mathsf{spawn}(s'',p')\cons h,p,s}_{\varphi} 
\comp \ttuple{h',p',s'}_{\varphi'}
& \looparrowright & \ttuple{\mathsf{spawn}(s'',p')\cons h,p,s}_{\varphi}
\comp \ttuple{h',p',s'}_{\varphi'\cup\{\mathsf{\tau},\mathsf{sp}\}} \\[2ex]

(\mathit{Send1}) & \ttuple{\mathsf{send}(s'',p',\ell)\cons h,p,s}_{\varphi} 
\comp (p,p',\{\ell,v\})
& \looparrowright & \ttuple{h,p,s''}_{\varphi} \\
(\mathit{Send2}) & \ttuple{\mathsf{send}(s'',p',\ell)\cons h,p,s}_{\varphi} 
\comp \ttuple{h',p',s'}_{\varphi'}
& \looparrowright & \ttuple{\mathsf{send}(s'',p',\ell)\cons h,p,s}_{\varphi}
\comp \ttuple{h',p',s'}_{\varphi'\cup\{\tau,\ell\}	} \\
\end{array}
\]
\caption{Controlled backward semantics} \label{fig:controlled-semantics}
\end{figure}
The controlled backward semantics is given by a transition
relation, $\looparrowright$, that can be found in 
Figure~\ref{fig:controlled-semantics}. Here, configurations
have the form $\ttuple{h,p,s}_\varphi$, where $h$ is a history,
$p$ a pid, $s$ an state, and $\varphi$ is a (possibly empty)
set of rollback requests.
Given a configuration $\tuple{h,p,s}$ and a checkpoint
identifier $\tau$, a rollback starts with an initial
configuration of the form $\ttuple{h,p,s}_{\{\tau\}}$.
In the following, in order to simplify the reduction 
rules, we consider that our semantics satisfies the 
following structural equivalence:
\[
(\mathit{SC}) ~~~ 
\ttuple{h,p,s}_\emptyset ~\equiv ~\tuple{h,p,s}
\]
Let us briefly explain the rules:\footnote{We assume a 
determinizing convention where a rule applies only 
if no earlier rule applies.}
\begin{itemize}
\item Rule \emph{Par} is identical to that in
the reversible semantics.
%, while rule \emph{Resume}
%simply terminates a rollback whenever the set of
%rollback requests is empty. 

\item Rule \emph{Seq} just updates the current state
and keep going backwards.

\item Rule \emph{Check} recovers the saved state and
removes $\tau$ from the current set of requests $\varphi$ 
(assuming $\tau\in\varphi$).

\item Rule \emph{SP} removes a request $\mathsf{sp}$ from
the current set $\varphi$ (if $\mathsf{sp}\in\varphi$) and
the configuration is in its initial state (i.e., with an
empty history). If the derivation is well defined, 
$\varphi$ will be empty after the step.

\item Rule \emph{Receive} just puts a received message back 
on the network and removes the request $\ell$ from $\varphi$
(assuming $\ell\in\varphi$).

\item Rules \emph{Spawn1} and \emph{Spawn2} deal with 
undoing a spawn. If the spawned process is in its initial state
(i.e., has an empty history), it is deleted and the element is
removed from the history. Otherwise, rule \emph{Spawn2} puts
the spawned process in rollback mode and adds the
requests $\tau$ and $\mathsf{sp}$ to its (possibly empty)
current set of rollback requests $\varphi'$.

\item Rules \emph{Send1} and \emph{Send2} take care of 
undoing the sending of a message. If the message is on the network,
the first rule deletes it and removes the corresponding element
from the history. Otherwise, rule \emph{Send2} propagates the
rollback mode to the receiver of the message adding the
requests $\tau$ and $\ell$ to its (possibly empty)
set of rollback requests $\varphi'$.
\end{itemize}
Note that we do not have a specific rule to end a rollback
and resume the forward computation (with the rules of 
Figure~\ref{fig:reversible-semantics-forward}) 
but it actually terminates 
when the set of rollbacks is empty by applying the 
structural equivalence above (\emph{SC}).
%
%We let $\looparrowright$ denote the union of transition
%relations
%$\rh$ and $\leadsto$ (instead of $\lh$).
%
%Proving that the controlled reversible semantics 
%modeled by $\looparrowright$ is 
%sound is straightforward and can be done as in
%\cite[Theorem~25]{LNPV18jlamp}. Here, 
In the following, we let
$\rolldel(S)$ denote the system that results from $S$
by removing all ongoing rollbacks. Furthermore, 
a system is \emph{initial} under the %controlled
reversible semantics if it is composed by a single 
process with an empty history. % and set of active rollbacks. 
A system $S$ is \emph{reachable} 
% under the controlled reversible semantics 
if there exist an initial system 
$S_0$ and a derivation $S_0 \rlh^\ast S$.

\begin{theorem} \label{th:rolldel-sound}
  Let $S$ be a reachable system under the %controlled
  reversible semantics. If $S \looparrowright^\ast S'$,
  then $\rolldel(S) \rlh^\ast \rolldel(S')$.
\end{theorem}

\begin{proof}
  The proof is perfectly analogous to that of
  Theorem~25 in \cite{LNPV18jlamp}. \qed
\end{proof}
%
%Observe that this is similar to the propagation of the
%rollback mode in our semantics, as we will see below.
% (check the second rules of 
%$\overline{\mathit{Send}}$ and $\overline{\mathit{Spawn}}$
%in Figure~\ref{fig:rollback-semantics-backward}).
%
In the following, we assume that 
%derivations are well-defined and, moreover, 
a process history always contain the complete sequence
of elements which are needed 
to perform an eventual rollback (this is an easy
consequence of the notion of well-defined derivation and the
definition of the rollback recovery semantics.
%rules in Figures~\ref{fig:rollback-semantics-core}
%and \ref{fig:rollback-semantics-commit}). 

The soundness of the forward rules 
(Figure~\ref{fig:rollback-semantics-core}) is 
straightforward, since every step has a direct counterpart
under either the standard or the (forward) reversible 
semantics. 
Now, we focus on the rollback rules
(Figure~\ref{fig:rollback-semantics-rollback}).

%We now introduce an equivalence relation between 
%our rollback recovery configurations and the 
%configurations of the reversible semantics:

\begin{definition}
  Let $S_{rr}$ be a system of the rollback recovery
  semantics and $S_r$ a system of the reversible semantics.
  We define an equivalence relation ``$\:\approx$'' as follows:
  $S_{rr}\approx S_r$ if $\sql S_{rr}\sqr = \sqll S_r \sqrr$, 
  where the auxiliary functions $\sql\cdot\sqr$ and $\sqll\cdot\sqrr$
  are defined as follows:
  \[
  \begin{array}{rcl}
  \sql S_1 \comp S_2 \sqr & = & \sql S_1 \sqr \comp \sql S_2 \sqr\\
  \sql (\cC,p,p',\{\ell,v\})\sqr & = & (p,p',\{\ell,v\})\\
  \sql \tuplefb{p,p',v}\sqr & = & \epsilon \\
%  ~~~\mbox{(i.e., it is removed from the system)}\\
  \sql \tuple{\Delta,p,s} \sqr & = & \tuple{\widetilde{\Delta},p,s}\\
  \sql \tuple{\Delta,p,s}^{\tau,p_\tau,p_r,L,P} \sqr & = & \tuple{\widetilde{\Delta},p,s}
  \end{array}
  \]
  where
  \[
  \widetilde{\Delta} = \left\{\begin{array}{ll}
    \mathit{send}(p,\ell)\cons\widetilde{\Delta'} & \mbox{if}~\Delta = \mathit{send}(p,\ell)\cons\Delta'\\
    \mathit{rec}(p,p',\{\ell,v\})\cons\widetilde{\Delta'} & \mbox{if}~\Delta = \mathit{rec}(\cC,p,p',\{\ell,v\})\cons\Delta'\\
    \mathit{spawn}(p)\cons\widetilde{\Delta'} & \mbox{if}~\Delta = \mathit{spawn}(p)\cons\Delta'\\
    \mathit{check}(\tau,s)\cons\widetilde{\Delta'} & \mbox{if}~\Delta = \mathit{check}(\tau,s)\cons\Delta'\\
    \widetilde{\Delta'} & \mbox{if}~\Delta = \overline{\mathit{check}}(\tau,s)\cons\Delta'\\
  \end{array}\right.
  \]
  and
  \[
  \begin{array}{rcl}
  \sqll S_1 \comp S_2 \sqrr & = & \sqll S_1 \sqrr \comp \sqll S_2 \sqrr\\
  \sqll (p,p',\{\ell,v\})\sqrr & = & (p,p',\{\ell,v\})\\
  \sqll \tuple{h,p,s} \sqrr & = & \tuple{\overline{h},p,s}
  \end{array}
  \]
  where
  \[
  \overline{h} = \left\{\begin{array}{ll}
    \overline{h'} & \mbox{if}~ h = \mathsf{seq}(s)\cons h'\\
    \mathit{send}(p,\ell)\cons\overline{h'} & \mbox{if}~h = \mathsf{send}(s,p,\ell)\cons h'\\
    \mathit{rec}(p,p',\{\ell,v\})\cons\overline{h'} & \mbox{if}~h = \mathsf{rec}(s,p,p',\{\ell,v\})\cons h'\\
    \mathit{spawn}(p)\cons\overline{h'} & \mbox{if}~h = \mathsf{spawn}(s,p)\cons h'\\
   \mathit{check}(\tau,s)\cons\overline{h'} & \mbox{if}~h = \mathsf{check}(\tau,s)\cons h'\\
  \end{array}\right.
  \]
\end{definition}
Now, we prove that, for every rollback derivation under 
the rollback recovery semantics, there exists an
equivalent rollback derivation under the controlled
backward semantics, 
%(Figure~\ref{fig:controlled-semantics}), 
and vice versa.
This will allow us to state the soundness of the
rollback recovery semantics
(together with Theorem~\ref{th:rolldel-sound}).
In the following, we only consider the case
$s\oplus s_\tau = s_\tau$ and omit sequential
steps in process histories for simplicity.

\begin{lemma} \label{lemma:rollback}
  Let $S_0 = \tuple{\Delta,p,s}\comp S_1$ 
  be a system of the
  rollback recovery semantics, where we assume that
  there is no ongoing rollback, and 
   $S'_0 = \tuple{h,p,s}\comp S'_1$ be a system
  of reversible semantics 
  such that $S_0\approx S'_0$.
  Then, we have a derivation $d$ of the form
  \[
  S_0 \hoo 
  \tupleb{\Delta',p,s_\tau}^{\tau,p_\tau,p_r,L,P}\comp S_1
  \hoo^\ast 
  \tupleb{\Delta',p,s_\tau}^{\tau,p_\tau,p_r,\emptyset,\emptyset}
  \comp S_2
  %\hoo^\ast %\tuple{\Delta',p,s_\tau}\comp S_3
  \]
  under the rollback recovery semantics
  iff there exists a derivation $d'$ of the form
  \[
  \ttuple{h,p,s}_{\{\tau\}} \comp S'_1
  \looparrowright^\ast \ttuple{h',p,s_\tau}_{\emptyset}\comp S'_2
  \] 
  under the controlled backward semantics
  such that %$\tuple{\Delta',p,s_\tau}\comp S_2
  $\tupleb{\Delta',p,s_\tau}^{\tau,p_\tau,p_r,\emptyset,\emptyset}
  \comp S_2 \approx \tuple{h',p,s_\tau}\comp S'_2$.
\end{lemma}

\begin{proof}
  We prove the ``if'' direction. The ``only if'' 
  will easily follow since the rollback rules in
  Figure~\ref{fig:rollback-semantics-rollback} are
  deterministic by definition. 
%  Furthermore, we assume a user-defined
%  rollback for simplicity. A rollback associated to a
%  forced checkpoint would proceed analogously by 
%  adding an additional element $\mathsf{check}(\tau,s)$
%  either to the end of the history (the case of 
%  a forced checkpoint associated to a spawned process)
%  or just before receiving a message (the case of
%  a forced checkpoint associated to message sending).
  %
  We prove the claim by induction on the \emph{depth}
  $D$ of the derivation, i.e., on the length of the maximum
  \emph{chain} of process dependencies (via either 
  $\mathsf{spawn}(s,p)$ or $\mathsf{send}(s,p,\ell)$) 
  in the derivation.
  
  Base case ($D=0$). In this case, $h$ can only have 
  occurrences of $\mathsf{rec}$ and $\mathsf{check}$ 
  (recall that we are ignoring sequential steps).
  Therefore, $d'$ performs a number of steps with rule
  \emph{Receive}, followed by a final step with rule
  \emph{Check}, then recovering the state in
  $\mathsf{check}(\tau,s)$. Trivially, the rollback
  rules achieve the same result in one step by applying either
  rule \emph{Rollback} or \emph{Roll1}, depending on
  whether it is a user-defined rollback or a rollback
  propagated from other process because of a forced
  checkpoint.
%  and \emph{Resume1} (if it is a user-defined rollback), 
%  or rules \emph{Roll1}, \emph{Resume3}, and \emph{Resume4}.
  
  Consider now the inductive case ($D>0$).
  Assume that $h$ has the form
  $h = a_1\cons\ldots a_n\cons
  \mathsf{check}(\tau,s_\tau)\cons h'$. 
  We make a case distinction on each element $a_i$
  in the history:
  \begin{itemize}
  \item $a_i = \mathsf{rec}(s',p,p',\{\ell,v\})$. This case
  is similar to the base case, since each element of this
  form will be undone by putting the message back on the
  network (rule \emph{Receive} in 
  Figure~\ref{fig:controlled-semantics}), similarly
  to either rule \emph{Rollback} or \emph{Roll1} in 
  Figure~\ref{fig:rollback-semantics-rollback}.
  
  \item $a_i = \mathsf{spawn}(s',p')$. In this case,
  we know that the history of process $p'$ must start
  with an element of the form $\mathsf{check}(\tau,\bot)$
  since $S_0 \approx S'_0$.\footnote{Originally, these
  forced checkpoints are not included in the controlled
  backward semantics (since they are not needed there).
  Here, however, we consider that these redundant elements have
  been added following the structure of $S_0$. 
  Note that it does
  not change the semantics since rule \emph{Check}
  (Figure~\ref{fig:controlled-semantics}) will eventually
  remove them anyway.} Since the history of $p'$ is never
  empty (it will contain, at least, one element:
  $\mathsf{check}(\tau,\bot)$), the controlled backward semantics
  will first apply rule \emph{Spawn2} to propagate the rollback
  to $p'$. Then, the controlled backward semantics
  will reduce $p'$ until its history is empty and 
  rule \emph{Spawn1} removes it.
  We have a similar behavior in the rollback recovery
  semantics by either applying rule \emph{Rollback}
  or \emph{Roll1} first, so that the rollback is propagated
  to process $p'$, and eventually applying rule \emph{Resume2}
  (which removes process $p'$) and rule
  \emph{Undo-dep1} (which removes the process dependency
  from the set $P$). Hence, the claim follows by 
  applying the inductive hypothesis
  to the rollback derivation for $p'$ since the depth
  of its derivation is strictly smaller.
  
  \item $a_i = \mathsf{send}(s',p',\ell)$. Here, we
  distinguish two cases. If this is not the \emph{oldest} 
  dependency with process $p'$ (i.e., there exists
  $a_j = \mathsf{spawn}(s'',p')$ or 
  $a_j = \mathsf{send}(s'',p',\ell')$ for some $j>i$),
  then it is ignored.
  This is safe since we know that a rollback to an
  earlier state will be introduced by the oldest 
  dependency. A similar behavior is obtained
  in the rollback recovery semantics by applying
  rule \emph{Roll3}. Otherwise (i.e., it is the oldest 
  dependency with process $p'$), we
  proceed similarly to the previous case:
  \begin{itemize}
  \item If the message has not been yet received by
  process $p'$, rule \emph{Send1} removes it from the
  network and we are done. A similar effect is achieved in
  the rollback recovery semantics by applying rule 
  $\mathit{Roll2}$, $\mathit{Resume3}$, 
  and $\mathit{Resume4}$.
  \item Otherwise, if the message has been received by
  process $p'$, we know that the history 
  of $p'$ must contain an
  element of the form $\mathsf{check}(\tau,s'')$ just
  before the element $\mathsf{rec}(s_r,p,p',\{\ell,v\})$
  since $S_0 \approx S'_0$. Hence, the controlled 
  backward semantics proceeds by applying rule \emph{Send2},
  then reducing $p'$ up to the removal of the
  element $\mathsf{check}(\tau,s'')$ and, then, 
  applying rule \emph{Send1} to remove the message
  from the network.
  In the rollback recovery semantics, we achieve a similar
  result by first applying either rule \emph{Rollback}
  or \emph{Roll1}, which propagates the rollback to process
  $p'$, and eventually applying rule \emph{Undo-send}
  to remove the message from the network. 
  As before, the claim then follows by applying the
  inductive hypothesis to
  the rollback derivation for $p'$. \qed
  \end{itemize}
  \end{itemize}
\end{proof}
The soundness of the rollback recovery semantics can
now be stated as follows. Here, we let $\to^+$ denote
the reflexive closure of relation $\to$.

\begin{theorem} \label{th:sound2}
  Let $d$ be a well-defined derivation. Then, every 
  forward step $S_1 \hoo S_2$ in $d$ can be projected to one
  or zero steps under the standard semantics:
  $sta(S_1) \boo^+ sta(S_2)$. Moreover, for every
  rollback derivation $S_1 \hoo^\ast S_2$ in $d$ using the rules 
  in Figure~\ref{fig:rollback-semantics-rollback}
  with $S_1\approx S'_1$, 
  there exists an equivalent derivation 
  $\mathsf{rolldel}(S'_1) \looparrowright \mathsf{rolldel}(S'_2)$
  under the reversible semantics
  such that $S_2 \approx S'_2$.
\end{theorem}

\begin{proof}
The first claim on the forward steps is a direct
consequence of Theorem~\ref{th:sound1}. 
We note that the Commit rules 
(Figure~\ref{fig:rollback-semantics-commit})
are considered \emph{forward} steps too. However,
they do not modify the projected system. This is why
we consider the reflexive closure in 
$sta(S_1) \boo^+ sta(S_2)$, since in this case we have
$sta(S_1) = sta(S_2)$.
The second claim about rollback derivations is an
easy consequence of Lemma~\ref{lemma:rollback} and
Theorem~\ref{th:rolldel-sound}.
\qed
\end{proof}

%%%%%%%%%%%%%%%%%%%%%%%%%%%%%%%%%%%%%%%%%%%%%%%%%%%%%%%%%%
\section{Discussion}\label{sec:relwork}

Our work shares some similarities with \cite{FV05}, where a new
programming model for globally consistent checkpoints is
introduced. However, we aim at extending an existing language
(like Erlang) rather than defining a new one. 
There is also a relation with \cite{SKM17}, which
presents a hybrid model combining message-passing concurrency 
and software transactional memory. 
However, the underlying language is different and, moreover, 
their transactions cannot include process spawning 
(which is delayed after the transition terminates).

A closely related approach is \cite{MTY23}, which 
introduces a rollback recovery
strategy for \emph{session-based programming}, where
some primitives for rollback recovery are defined. 
However, they consider a different setting 
(a variant of $\pi$-calculus) and the number of parties 
is fixed (no dynamic process spawning); furthermore, 
nested checkpoints are not allowed.  
%Furthermore, the checkpoints of \cite{MTY23} are
%not automatically propagated to other causally consistent
%processes (as our forced checkpoints); rather, they introduce a
%\emph{compliance check} at the type level to prevent 
%undesired situations.
Also, 
%Finally, Fabbretti, Lanese and Stefani 
\cite{FLS23} presents a
calculus to formally model distributed systems subject to crash 
failures, where recovery mechanisms can be encoded by
a small set of primitives. As in the previous case, a
variant of $\pi$-calculus is considered. 
%This work can be seen
%as a reworking and extension of the previous work by
%Francalanza and Hennessy \cite{FH08}. 
%Here, a variant of $\pi$-calculus is considered. 
Furthermore,
the authors focus on crash recovery without relying on
a form of checkpointing, in contrast to our approach.

The closest approach is that of \cite{Vid23facs}, where 
causal consistent rollback recovery for message-passing
concurrent programs is considered. However, there are significant
differences with our approach. First, \cite{Vid23facs} defines
a rollback procedure based on a reversible semantics, which means that
a process must save the state in \emph{every} step. Moreover,
a rollback implies undoing \emph{all} the actions of a process
in a stepwise manner (a consequence of the fact that the reversible
semantics was originally introduced for reversible debugging
in \cite{LNPV18jlamp}). Furthermore, the operational semantics
in \cite{Vid23facs} is not fully asynchronous. All in all, it
represents an interesting theoretical result but cannot be
used as a basis for a practical implementation.

In contrast, in this work we have 
designed a rollback recovery strategy
for a message-passing concurrent language 
%(like Erlang \cite{erlang}) 
that is purely asynchronous and does not need 
a central coordination. Therefore, it represents a good
foundation for the development of a practical
implementation of rollback recovery based on a
source-to-source program instrumentation.
As future work, we plan to develop a proof-of-concept
implementation of the proposed scheme and to further
study the properties of the rollback 
recovery semantics (including efficiency issues). 
In particular, we will 
experimentally evaluate the behavior of the rollback
recovery strategy on larger, more realistic examples.

\subsubsection*{Acknowledgements.}

The author would like to thank Ivan Lanese 
and Adri\'an Palacios for their useful remarks 
and discussions on a preliminary version of
this work. 
I would also like to thank the anonymous reviewers 
for their suggestions to improve this paper.

%\bibliography{biblio}

\begin{thebibliography}{10}
\providecommand{\url}[1]{\texttt{#1}}
\providecommand{\urlprefix}{URL }
\providecommand{\doi}[1]{https://doi.org/#1}

\bibitem{EAWJ02}
Elnozahy, E.N., Alvisi, L., Wang, Y., Johnson, D.B.: A survey of
  rollback-recovery protocols in message-passing systems. {ACM} Comput. Surv.
  \textbf{34}(3),  375--408 (2002)

\bibitem{erlang}
URL: \url{https://www.erlang.org/} (2021)

\bibitem{FLS23}
Fabbretti, G., Lanese, I., Stefani, J.B.: A behavioral theory for crash
  failures and erlang-style recoveries in distributed systems. Tech. Rep.
  RR-9511, INRIA (2023), \url{https://hal.science/hal-04123758}

\bibitem{FV05}
Field, J., Varela, C.A.: Transactors: a programming model for maintaining
  globally consistent distributed state in unreliable environments. In:
  Palsberg, J., Abadi, M. (eds.) Proceedings of the 32nd {ACM} {SIGPLAN-SIGACT}
  Symposium on Principles of Programming Languages ({POPL} 2005). pp. 195--208.
  {ACM} (2005)

\bibitem{FMT18}
Francalanza, A., Mezzina, C.A., Tuosto, E.: Reversible choreographies via
  monitoring in erlang. In: Bonomi, S., Rivi{\`{e}}re, E. (eds.) Proceedings of
  the 18th {IFIP} {WG} 6.1 International Conference on Distributed Applications
  and Interoperable Systems ({DAIS} 2018), held as part of DisCoTec 2018.
  Lecture Notes in Computer Science, vol. 10853, pp. 75--92. Springer (2018).
  \doi{10.1007/978-3-319-93767-0\_6}

\bibitem{HBS73}
Hewitt, C., Bishop, P.B., Steiger, R.: A universal modular {ACTOR} formalism
  for artificial intelligence. In: Nilsson, N.J. (ed.) Proceedings of the 3rd
  International Joint Conference on Artificial Intelligence. pp. 235--245.
  William Kaufmann (1973),
  \url{http://ijcai.org/Proceedings/73/Papers/027B.pdf}

\bibitem{HYC08}
Honda, K., Yoshida, N., Carbone, M.: Multiparty asynchronous session types. In:
  Necula, G.C., Wadler, P. (eds.) Proceedings of the 35th {ACM}
  {SIGPLAN-SIGACT} Symposium on Principles of Programming Languages ({POPL}
  2008). pp. 273--284. {ACM} (2008). \doi{10.1145/1328438.1328472}

\bibitem{Lam78}
Lamport, L.: Time, clocks, and the ordering of events in a distributed system.
  Commun.\ {ACM}  \textbf{21}(7),  558--565 (1978). \doi{10.1145/359545.359563}

\bibitem{LM20}
Lanese, I., Medic, D.: A general approach to derive uncontrolled reversible
  semantics. In: Konnov, I., Kov{\'{a}}cs, L. (eds.) 31st International
  Conference on Concurrency Theory, {CONCUR} 2020. LIPIcs, vol.~171, pp.
  33:1--33:24. Schloss Dagstuhl - Leibniz-Zentrum f{\"{u}}r Informatik (2020).
  \doi{10.4230/LIPIcs.CONCUR.2020.33}

\bibitem{LNPV18jlamp}
Lanese, I., Nishida, N., Palacios, A., Vidal, G.: A theory of reversibility for
  {E}rlang. Journal of Logical and Algebraic Methods in Programming
  \textbf{100},  71--97 (2018). \doi{10.1016/j.jlamp.2018.06.004}

\bibitem{LPV19}
Lanese, I., Palacios, A., Vidal, G.: Causal-consistent replay debugging for
  message passing programs. In: P{\'{e}}rez, J.A., Yoshida, N. (eds.)
  Proceedings of the 39th {IFIP} {WG} 6.1 International Conference on Formal
  Techniques for Distributed Objects, Components, and Systems ({FORTE} 2019).
  Lecture Notes in Computer Science, vol. 11535, pp. 167--184. Springer (2019).
  \doi{10.1007/978-3-030-21759-4\_10}

\bibitem{LPV21}
Lanese, I., Palacios, A., Vidal, G.: Causal-consistent replay reversible
  semantics for message passing concurrent programs. Fundam. Informaticae
  \textbf{178}(3),  229--266 (2021). \doi{10.3233/FI-2021-2005}

\bibitem{LSZ19}
Lanese, I., Sangiorgi, D., Zavattaro, G.: Playing with bisimulation in
  {E}rlang. In: Boreale, M., Corradini, F., Loreti, M., Pugliese, R. (eds.)
  Models, Languages, and Tools for Concurrent and Distributed Programming --
  Essays Dedicated to Rocco De Nicola on the Occasion of His 65th Birthday.
  Lecture Notes in Computer Science, vol. 11665, pp. 71--91. Springer (2019).
  \doi{10.1007/978-3-030-21485-2\_6}

\bibitem{MTY23}
Mezzina, C.A., Tiezzi, F., Yoshida, N.: Rollback recovery in session-based
  programming. In: Jongmans, S., Lopes, A. (eds.) Proceedings of the 25th
  {IFIP} {WG} 6.1 International Conference on Coordination Models and
  Languages, {COORDINATION} 2023. Lecture Notes in Computer Science, vol.
  13908, pp. 195--213. Springer (2023). \doi{10.1007/978-3-031-35361-1\_11}

\bibitem{SKM17}
Swalens, J., Koster, J.D., Meuter, W.D.: Transactional actors: communication in
  transactions. In: Jannesari, A., de~Oliveira~Castro, P., Sato, Y., Mattson,
  T. (eds.) Proceedings of the 4th {ACM} {SIGPLAN} International Workshop on
  Software Engineering for Parallel Systems, SEPS\@SPLASH 2017. pp. 31--41.
  {ACM} (2017). \doi{10.1145/3141865.3141866}

\bibitem{Vid23facs}
Vidal, G.: From reversible computation to checkpoint-based rollback recovery
  for message-passing concurrent programs. In: C\'{a}mara, J., Jongmans, S.S.
  (eds.) Formal Aspects of Component Software - 19th International Conference,
  {FACS} 2023, Virtual Event, October 19-20, 2023, Proceedings. Lecture Notes
  in Computer Science, Springer (2023), to appear (see
  https://arxiv.org/abs/2309.04873)

\end{thebibliography}
\bibliographystyle{splncs04}

\end{document}